\documentclass[aps,prb,twocolumn,groupedaddress,showpacs]{revtex4}

\usepackage[dvips]{graphicx}
\usepackage{amsmath}
\usepackage{amstext}
\usepackage{graphics}
\usepackage{exscale}
\usepackage{epsfig}
\usepackage[T1]{fontenc}
\usepackage{bm}
\usepackage{bbm}

\usepackage{latexsym}
\renewcommand{\epsilon}{\varepsilon}

\newcommand{\integral}[3]{\!\int\limits_{#2}^{#3}\!\!{\rm d}#1\;}

\newcommand{\expval}[2]{ \langle  #1 #2\ \!\! \rangle}

\newcommand{\elcre}[2]{ c^{\dagger}_{#1,#2}}
\newcommand{\elann}[2]{ c_{#1,#2}}

\newcommand{\vct}[1]{\bm #1}
\newcommand{\vk}{{\bm k}}

\newcommand{\ord}[1]{{\cal O}(#1)}

\newcommand{\thGf}{{\cal G}}

\newcommand{\hc}{\mathrm{h.c.}}

\begin{document}

\title{Field dependent quasiparticles in the infinite dimensional Hubbard model}
\author{J. Bauer and A.C. Hewson}
\affiliation{Department of Mathematics, Imperial College, London SW7 2AZ,
  United Kingdom}
\date{\today} 
\begin{abstract}
We present dynamical mean field theory (DMFT) results for the local spectral
densities of the one- and two-particle response functions for the infinite
dimensional Hubbard model in a magnetic field. We look at the different
regimes corresponding to half-filling, near half-filling 
and well away from half-filling, for intermediate and strong values of the local
interaction $U$. The low energy results are analyzed in terms of
quasiparticles with field dependent parameters. The renormalized parameters
are determined by two different methods, both based on numerical
renormalization group (NRG) calculations, and we find good agreement. Away
from half-filling the quasiparticle weights, 
$z_\sigma(H)$, differ according to the spin type $\sigma=\uparrow$ or
$\sigma=\downarrow$. Using the renormalized parameters, we show that DMFT-NRG
results for the local longitudinal and transverse dynamic spin susceptibilities in an
arbitrary field can be understood in terms of repeated scattering of these
quasiparticles. We also check Luttinger's theorem for the Hubbard model and
find it to be satisfied in all parameter regimes and for all values of the
magnetic field.   
 \end{abstract}
%\PACS{ 72.15.Qm\sep 75.20Hr\sep 73.21.La}
%72.10.F Kondo transport
% 73.61 Quantum dots conductivity
% 71.10.A Fermi liquid theory
% 11.10.G Renormalization, field theory

\maketitle

\section{Introduction}

A feature of strongly correlated electron systems, such as heavy fermions,
is their sensitivity  to an applied magnetic field, which makes  a magnetic
field a useful experimental probe of strong correlation behavior.  A
manifestation of this sensitivity is the very  large paramagnetic
susceptibility observed in these systems. In terms of Fermi liquid theory, the
large paramagnetic susceptibility can be interpreted as due to quasiparticles
with exceptionally large effective masses. These large effective masses arise
from the scattering of the electrons with the enhanced spin 
fluctuations induced by the strong local Coulomb interactions.
An applied magnetic field suppresses the spin fluctuations causing a reduction
in the effective masses, which can be seen experimentally in de Haas-van
Alphen measurements \cite{JRCTDF87,GHTYF99}. Not only do 
the effective masses depend on the magnetic field, they may also  differ for
the spin up and spin down electrons \cite{AUAO93,KSWA95}. Another feature that reflects
the enhanced sensitivity to an applied field is metamagnetic behavior, where
the spin susceptibility $\chi(H)$ in a finite field $H$ increases with the
field strength such that $d\chi(H)/dH>0$, which has been observed in some heavy
fermion compounds \cite{MCCSRC00}. This can be understood in terms of an increase in
the effective mass for larger fields opposite to the effect described above. It
is related to the fact that strong magnetic fields can induce localization in
narrow conduction bands as predicted theoretically \cite{Vol84,LGK94,JC00}.
This has been observed experimentally, for instance,  in quasi-two dimensional
organic conductors \cite{KIMK04}. 

A lattice model which can mimic many of these effects is the single band Hubbard model.
It has played a similar role for lattice models as the Anderson model for
impurity models, being the simplest model of its 
type, where the interplay of kinetic energy and strong local interactions can
be studied. Here we are interested in studying the magnetic response of
this model for different interactions and fillings.
The calculations are based on the dynamical mean field theory (DMFT)
with the numerical renormalization group (NRG) to solve the effective impurity
problem. We present results for the local spectral densities and spin
dynamics in parameter regimes at half filling
and for finite hole doping, where different responses to the magnetic field
can be observed. 

We interpret the low energy results in terms of
quasiparticles which are characterized by field dependent renormalized
parameters. The approach is similar to that used earlier \cite{HBK06,BH07apre}
for the Anderson model in a magnetic field.
We deduce the renormalized parameters by two different methods based on the
NRG calculations. 
These parameters can be used to define a free
quasiparticle density of states, which gives the asymptotically exact spectral
behavior for low energy. The effects of the interactions between these quasiparticles can be
taken into account using a renormalized perturbation theory (RPT) \cite{Hew93,Hew01}.
It was shown earlier that a very good description of the $T=0$ spin and charge dynamics
for the Anderson model in the Kondo regime can be obtained  by summing the RPT
diagrams for repeated quasiparticle scattering \cite{Hew06}. 
Here we extend these RPT calculations for the spin dynamics to
the lattice case and show that we can also understand the DMFT-NRG for the local
dynamic spin susceptibilities in terms of quasiparticle scattering.

\section{Dynamical mean field approach and renormalized parameters}
The Hamiltonian for the Hubbard model in a magnetic field in the grand
canonical formulation is given by
\begin{equation}
H_{\mu}=\sum_{i,j,\sigma}(t_{ij}\elcre {i}{\sigma}\elann
{j}{\sigma}+\hc)-\sum_{i\sigma}\mu_{\sigma}
n_{i\sigma}+U\sum_in_{i,\uparrow}n_{i,\downarrow}\label{hubm}, 
\end{equation}
where $t_{ij}$ are the hopping matrix elements between sites $i$ and $j$,
$U$ is the on-site interaction; $\mu_{\sigma}=\mu+\sigma h$, where $\mu$ is the
chemical potential of the interacting system, and the Zeeman splitting term
with external magnetic field $H$ is given by $h=g\mu_{\rm B} H/2$, where
$\mu_{\rm B}$ is the Bohr magneton. We are dealing with the one s-band Hubbard
model here, so no coupling of the field to angular momentum states has to be included.

From Dyson's equation, the Green's function $G_{{\vk},\sigma}(\omega)$ can
be expressed in the form,  
\begin{equation}
G_{{\vk},\sigma}(\omega)
=\frac{1}{\omega+\mu_\sigma-\Sigma_{\sigma}({\vk},\omega)-\epsilon({\vk})},  
\end{equation}
where $\Sigma_{\sigma}({\vk},\omega)$ is the proper self-energy,
and $\epsilon({\vk})=\sum_{\vk}e^{-{\vk}\cdot({\vct R}_i-{\vct R}_ j)}t_{ij}$.
The simplification that occurs for the model in the infinite dimensional limit
is that $\Sigma_{\sigma}({\vk},\omega)$  becomes
a function of $\omega$ only \cite{MV89,Mue89}. In this case the local Green's function  
$ G_{\sigma}^{\mathrm{loc}}(\omega)$ can be expressed in the form,
\begin{equation}
G_{\sigma}^{\mathrm{loc}}(\omega) =\sum_{\vk}G_{{\vk},\sigma}(\omega)=
\integral{\epsilon}{}{}\frac{D(\epsilon)}
{\omega+\mu_\sigma -\Sigma_{\sigma}(\omega)-\epsilon},
\label{gloc} 
\end{equation}
where $D(\epsilon)$ is the density of states for the non-interacting model
($U=0$). It is possible to convert this lattice problem into an effective
impurity problem \cite{GKKR96}. We introduce the dynamical Weiss field
$\thGf_{0,\sigma}^{-1}(\omega)$ and write the Green's function in the form
\begin{equation}
 G^{\mathrm{loc}}_{\sigma}(\omega) =
\frac{1}{ \thGf_{0,\sigma}^{-1}(\omega) -\Sigma_{\sigma}(\omega)},
\label{locgf}
\end{equation}
which is equivalent to 
\begin{equation}
  \thGf_{0,\sigma}^{-1}(\omega)=G_{\sigma}^{\mathrm{loc}}(\omega)^{-1}
  +\Sigma_{\sigma}(\omega).
\label{tgf}
\end{equation}
The Green's function  $ G_{\sigma}^{\mathrm{loc}}(\omega)$ can  be
identified with the Green's function
$ G_{\sigma}(\omega)$ of an effective Anderson model,  by
re-expressing $\thGf_{0,\sigma}^{-1}(\omega)$ as
\begin{equation}
\thGf_{0,\sigma}^{-1}(\omega)=\omega+\mu+\sigma h-K_{\sigma}(\omega),
\label{thgfK}
\end{equation}
so that
\begin{equation}
G_{\sigma}(\omega)=\frac{1}
    {\omega-\epsilon_{\mathrm{d}\sigma}-K_\sigma(\omega)-\Sigma_\sigma(\omega)},
\label{gfdmft}
\end{equation}
 with $\epsilon_{\mathrm{d}\sigma}=-\mu_\sigma$. The
function $K_\sigma(\omega)$ plays the role of a dynamical mean field
describing the effective medium surrounding the impurity. In the impurity case
in the wide band limit we have $K_{\sigma}(\omega)=-i\Delta$. Here, as can be seen from
equations (\ref{tgf}) and (\ref{locgf}), $K_\sigma(\omega)$ is a
function of the self-energy $\Sigma_\sigma(\omega)$, and hence depends on
$\sigma$.  As this self-energy is identified with the
impurity self-energy, which  will depend on  the form taken for $K_\sigma(\omega)$, it is
clear that this quantity has to be calculated self-consistently. Starting from
an initial form for $K_\sigma(\omega)$, $\Sigma_\sigma(\omega)$ is
calculated using an appropriate 'impurity solver' from which 
$ G_{\sigma}^{\mathrm{loc}}(\omega)$ can be calculated using equation
(\ref{gloc}), and a new result for $K_\sigma(\omega)$ from equations
(\ref{tgf}) and (\ref{thgfK}). This $K_\sigma(\omega)$ serves as an input for
the effective impurity problem and the process is repeated until it converges 
to give a self-consistent solution. These equations constitute the dynamical mean field
theory (DMFT), and further details can be found in the review article of
Georges et al. \cite{GKKR96}.
\par
We need to specify the density of states $D(\omega)$ of the non-interacting
infinite dimensional model, which is usually taken to be either for a
tight-binding hypercubic or Bethe lattice. Here we take the
 semi-elliptical  form corresponding to a Bethe lattice,
\begin{equation}
    D(\epsilon)=\frac{2}{\pi D^2}\sqrt{D^2-(\epsilon+ \mu_0)^2}
\label{dos}
 \end{equation}
where $2D$ is the band width, with $D=2t$ for the Hubbard model, and $\mu_0$
the chemical potential of the free electrons. We choose
this form, rather than the Gaussian density of states of the hypercubic lattice,
as it has a finite bandwidth.\par 

Before considering in detail the methods of solving these equations, we look at the
form of these equations in the low energy regime, where we can give them an
interpretation in terms of renormalized quasiparticles.
We assume that we can expand $\Sigma_\sigma(\omega)$ in powers of $\omega$ for
small $\omega$, and retain  terms to first order in $\omega$ only.
Substituting this expansion into the equation for the local Green's function gives
\begin{equation}
 G_{0,\sigma}^{\mathrm{loc}}(\omega) =z_\sigma\integral{\epsilon}{}{}\frac{D(\epsilon/z_{\sigma})}
{\omega+\tilde\mu_{0,\sigma}+\ord{\omega^2} -\epsilon},
\label{expg}
\end{equation}
where
\begin{equation} \tilde\mu_{0,\sigma}=z_{\sigma}(\mu_{\sigma}-\Sigma_{\sigma}(0)),\quad{\rm and}\quad
     z_{\sigma}=1/[1-\Sigma'_{\sigma}(0)].
\label{nrgqp}
\end{equation}
 We have assumed the Luttinger
result that the imaginary part  of the self-energy vanishes at $\omega=0$.  
As the Green's function in equation (\ref{expg}) has the same form as in 
the non-interacting system, apart from the weight factor 
 $z_\sigma$,
we can use it to define a  free quasiparticle propagator,  $\tilde
G_{0,\sigma}^{\mathrm{loc}}(\omega)$,
\begin{equation}
\tilde G_{0,\sigma}^{\mathrm{loc}}(\omega) =\integral{\epsilon}{}{}\frac{D(\epsilon/z_{\sigma})}
{\omega+\tilde\mu_{0,\sigma} -\epsilon}.
\label{gqploc}
\end{equation}
We then interpret $z_\sigma$ as the quasiparticle weight.
We will refer to the density of states  $\tilde \rho_{0,\sigma}(\omega)$
derived from this Green's function via $\tilde \rho_{0,\sigma}(\omega)=-{\rm
  Im}\tilde G_{0,\sigma}(\omega+i\delta)/\pi$  as the free quasiparticle
density of states (DOS). For the Bethe lattice, the quasiparticle DOS 
takes the form of a band with renormalized parameters,
\begin{equation}
   \tilde \rho_{0,\sigma}(\omega)=\frac{2}{\pi\tilde D_{\sigma}^2}\sqrt{\tilde 
   D_{\sigma}^2-(\omega+\tilde \mu_{0,\sigma})^2},
\label{rdos}
 \end{equation}
where $\tilde D_{\sigma}=z_{\sigma}D$. 
We can also define a quasiparticle occupation number 
 $ \tilde n^0_{\sigma}$ by integrating this density of states up to the Fermi level,
 \begin{equation}
   \tilde n^0_{\sigma}=\integral{\omega}{-\infty}{0}\tilde
   \rho_{0,\sigma}(\omega).
\label{qpocc}
 \end{equation}
With a generalization of Luttinger's theorem \cite{Lut60} for each spin
component it is possible to relate this free quasiparticle 
occupation number  $ \tilde n^0_{\sigma}$ to the expectation value of the 
occupation number $n_{\sigma}$  in the interacting system at $T=0$.
Using the quasiparticle density of states in equation (\ref{rdos}),
we can rewrite equation (\ref{qpocc}) as
 \begin{equation}
   \tilde n^0_{\sigma}=\integral{\epsilon}{-\infty}{\infty} D(\epsilon)
\theta(\mu_\sigma-\Sigma_\sigma(0)-\epsilon),
\label{nocc}
 \end{equation}
where $\theta(\epsilon)$ is the Heaviside step function and $D(\epsilon)$ as given
in equation (\ref{dos}). Assuming Luttinger's result for each spin
component, the right-hand side of equation (\ref{nocc}) is equal to
$n_{\sigma}$. We then have the result, 
 \begin{equation}
   \tilde n^0_{\sigma}= n_{\sigma},
\label{maglat}
 \end{equation}
that the  occupation for electrons of spin $\sigma$ is equal to
the number of  free quasiparticle of spin $\sigma$, as calculated from
equation (\ref{qpocc}). It should be noted that there is no simple
generalization of the $h=0$ result, $\mu-\mu_0=\Sigma(0)$, in the spin
polarized case.

To solve the DMFT equations we need an 'impurity solver',
and the most commonly used are the quantum Monte Carlo, the
exact diagonalization (ED) method and the NRG,  all of which have advantages
and disadvantages. Here we use the NRG approach as it is the most accurate method
for calculations at $T=0$ and for the
low energy excitations. There has been a DMFT study of the static properties 
of a half-filled  Hubbard model in a magnetic
field  using the ED method by Laloux et al. \cite{LGK94}. 
The focus of our paper here, however, is rather different so there is little overlap
with this earlier work but, where there is, we make comment and compare with their results. 
 
To evaluate the renormalized parameters,  $z_\sigma$ and $\mu_{0,\sigma}$,
which specify the form of the quasiparticle DOS, we  use two different
methods based on the NRG approach. The first method is a direct one, where
we use the NRG determined 
the self-energy $\Sigma_\sigma(\omega)$ and the chemical potential
 $\mu_\sigma$, and then substitute into  equation (\ref{nrgqp}) for
 $z_\sigma$ and $\mu_{0,\sigma}$. The second method is indirect, 
and makes no reference  to the self-energy. It is based on the quasiparticle
 interpretation
of the NRG low energy fixed point of the effective impurity.  To explain this
 approach we need to consider in a little detail how the NRG calculations are carried out.

In the NRG approach \cite{KWW80a} the conduction band is logarithmically discretized
and the model then converted into the form of a one dimensional tight
binding chain, coupled via an effective hybridization $V_\sigma$ to the impurity
at one end.  In this representation $K_\sigma(\omega)=|V_\sigma|^2
g_{0,\sigma}(\omega)$, where $g_{0,\sigma}(\omega)$ is the one-electron Green's
function for the first site of the isolated conduction electron chain.
As earlier, we expand the self-energy
$\Sigma_\sigma(\omega)$ to first order in $\omega$,
and then substitute the result  into  equation (\ref{gfdmft}).
We then define a free quasiparticle  propagator, $\tilde
G_{0,\sigma}(\omega)$, for the impurity site as
\begin{equation}
\tilde
G_{0,\sigma}(\omega)=\frac{1}{\omega-\tilde\epsilon_{\mathrm{d}\sigma}-|\tilde
  V_\sigma|^2g_{0,\sigma}(\omega)}, 
\label{qpgfdmft}
\end{equation}
where
\begin{equation}
\tilde\epsilon_{\mathrm{d}\sigma}=z_\sigma(\epsilon_{\mathrm{d}\sigma}+\Sigma_\sigma(0)),\quad
|\tilde V_\sigma|^2={z_\sigma}| V_\sigma|^2, 
\label{rp}
\end{equation}
In the DMFT approach we identify $\tilde G_{0,\sigma}(\omega)$ with the local
quasiparticle  Green's function for the lattice (\ref{gqploc}), 
\begin{equation}
  \tilde G_{0,\sigma}^{\mathrm{loc}}(\omega)=\tilde G_{0,\sigma}(\omega),
\end{equation}
which  specifies the form of $g_{0,\sigma}(\omega)$ in (\ref{qpgfdmft}) 
and yields $\tilde\mu_{0,\sigma}=-\tilde\epsilon_{\mathrm{d}\sigma}$.
By fitting the low energy single particle excitations found in the NRG results
to the poles of this Green's function, we can deduce  the parameters $\tilde\epsilon_{\mathrm{d}\sigma}$
and $\tilde V_\sigma$, as has been explained in earlier work \cite{HOM04}. The
quasiparticle weight $z_\sigma$ is then obtained from the relation
$z_\sigma=|\tilde V_\sigma/ V_\sigma|^2$ in equation (\ref{rp}), and
$\tilde\mu_{0,\sigma}$ from $\tilde\mu_{0,\sigma}=-\tilde\epsilon_{\mathrm{d}\sigma}$.

Using the DMFT-NRG approach we can calculate the spectral densities for the local
two-particle response functions, as well as single-particle ones. The main
interest here will be in the local longitudinal and transverse dynamic spin
susceptibilities, $\chi_l(\omega)$ and  $\chi_t(\omega)$. Having calculated
the renormalized parameters, which describe the free quasiparticles, we can
compare the  DMFT-NRG results for the dynamic susceptibilities with the
corresponding quantities calculated via a renormalized RPA-like
 treatment that takes account of repeated quasiparticle-quasihole
scattering. This approach has been described fully elsewhere for the Anderson impurity
model \cite{Hew06,BH07apre}. The calculations here proceed along similar lines
as for the effective impurity model.  The equation for the transverse susceptibility is 
\begin{equation}
\chi_{t}(\omega)={\tilde\chi_{\uparrow\downarrow}(\omega)
\over 1-\tilde
U_t(h)\tilde\chi_{\uparrow\downarrow}(\omega)},
\label{trrpt}
\end{equation}
where $\tilde\chi_{\uparrow\downarrow}(\omega)$ is transverse susceptibility
calculated using  the free quasiparticles density of states given in equation
(\ref{rdos}),
and $\tilde U_t(h)$ is the irreducible quasiparticle interaction in this
channel.  For the Anderson model it was possible to calculate $\tilde U_t(h)$
in terms of the renormalized  on-site interaction  $\tilde U$.
Though it is possible in the lattice case to calculate  $\tilde U$,
we have no way of deducing $\tilde U_t(h)$ from it since unlike in the
impurity case  we do not have an exact expression for $\chi_t(0)$ in terms of
$\tilde U$. We determine it to fit the DMFT-NRG result for ${\rm
 Re}\chi_t(\omega)$   at the single  point $\omega=0$.
The corresponding result for the longitudinal susceptibility $\chi_{l}(\omega)$
is
\begin{eqnarray}
&&\chi_{l}(\omega)=  \label{lrrpt} \\
&&{\tilde\chi_{\uparrow\uparrow}(\omega,h)
 +\tilde\chi_{\downarrow\downarrow}(\omega,h)+4\tilde
 U_l(h)\tilde\chi_{\uparrow\uparrow}(\omega,h)\tilde\chi_{\downarrow\downarrow}(\omega,h)\over 2(
 1-4\tilde
 U^2_l(h)\tilde\chi_{\uparrow\uparrow}(\omega,h)\tilde\chi_{\downarrow\downarrow}(\omega,h))},
\nonumber
 \end{eqnarray}
where the susceptibilities  $\tilde\chi_{\sigma\sigma}(\omega)$ 
are those for the free quasiparticles, and $\tilde U_l(h)$ is determined by
fitting the DMFT-NRG result for  ${\rm Re}\chi_l(0)$.

Having covered the basic theory, we are now in a position to survey
the results for the Hubbard model in different parameter regimes.

\section{Results at Half-filling}

We present the results at half-filling for three main parameters regimes where we find
qualitatively different behavior. The results in all cases will be for a
Bethe lattice with a band width  $W=2D=4$, setting $t=1$ as the energy
scale. The local spectral densities are calculated from equation 
(\ref{gloc}) using the NRG deduced self-energy \cite{BHP98}. 
In the evaluation of all NRG spectra we use the improved method
\cite{PPA06,WD06pre} based on the complete Anders-Schiller basis \cite{AS05}. 
In concentrating on the field induced polarization, we do not include
the possibility of antiferromagnetic ordering. The regimes are (a) a
relatively weak coupling 
regime where $U$ is smaller than the band width, (b) an intermediate coupling 
regime with $W<U<U_c$, where $U_c$ is the value at which a Mott-Hubbard gap
develops in the absence of a magnetic field ($U_c\approx 5.88$)\cite{Bul99}, and (c) a
strong coupling regime  with $U>U_c$. 

\subsection{Weakly correlated regime}
\begin{figure}[!htbp]
\centering
\includegraphics[width=0.45\textwidth]{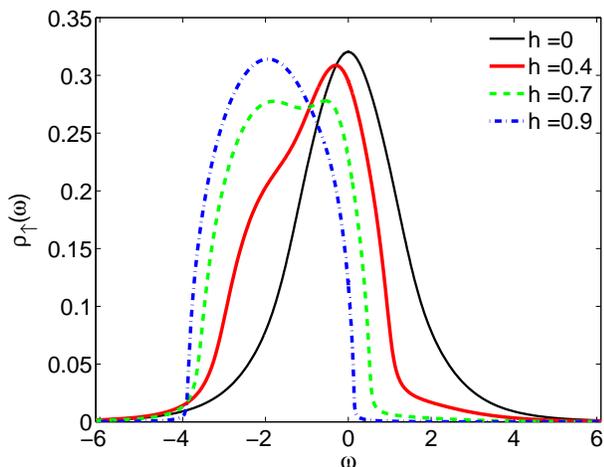}
\vspace*{-0.5cm}
\caption{(Color online) The local spectral density for the majority spin
  $\rho_\uparrow(\omega)$ for $U=2$ and various fields $h$. }  
\label{doswchdep}
\end{figure}

The first plot in figure \ref{doswchdep} gives the spectral densities for the
majority spin electrons $\rho_\uparrow(\omega)$ for various magnetic field values
in the weakly correlated regime, $U=2$. 

\begin{figure}[!htbp]
\centering
\includegraphics[width=0.45\textwidth]{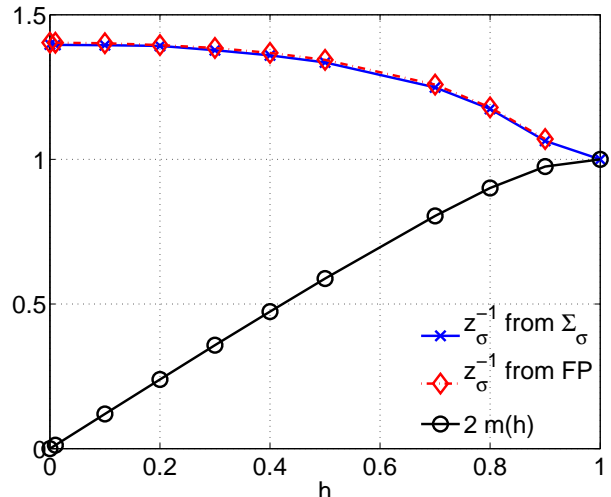}
\vspace*{-0.5cm}
\caption{(Color online) The inverse of the quasiparticle weight $z_\sigma(h)$ calculated
  from the impurity fixed point (FP) and directly from the self-energy and the
  magnetization $m(h)$ for $U=2$ and various fields $h$. }
\label{qpweightcompU2}
\end{figure}

\noindent
We can see clearly that, for increasing magnetic field, more and more spectral
weight is shifted to lower energies  (the opposite happens for the other spin
component, which is not displayed here). Above $h\simeq 1.0$ the system
is completely polarized, $2m=1$. This extreme high field limit corresponds to
a band insulator. There is a gap of the magnitude  $\Delta_g(h)=2h+U-W$  between the 
upper (minority) and lower (majority) bands, which both have the
semi-elliptical form  as for the non-interacting system  with $W=4$. At this point
dynamic renormalization effects have vanished. The inverse
of the quasiparticle weight  $z_\sigma(h)$, which corresponds to the
enhancement of the effective mass $m^*_{\sigma}(h)=m/z_{\sigma}(h)$,
is shown as a function of $h$ in figure \ref{qpweightcompU2}. 

\noindent
We calculated the values of  $z_\sigma(h)$ using the two methods described
earlier. At half filling we have $z_{\uparrow}(h)=z_{\downarrow}(h)$ and we
have plotted the average of the values for $\sigma={\uparrow}$ and
$\sigma={\downarrow}$, which is compared for the two methods. The deviation
for the values for the different spins is only due to small numerically
inaccuracies and is less than 2\%. The method based on analyzing the
excitations of the impurity fixed point (FP) is only applicable in the metallic
regime and when the system is not completely polarized.   
From these results shown in figure \ref{qpweightcompU2} it
can be seen that the two sets of values are in good agreement. 
The values of $z_\sigma(h)$  increase from about $0.75$ to $1$, which
corresponds to a progressive ``de-renormalization'' 
of the quasiparticles with increasing field, as observed earlier for the impurity model
\cite{HBK06}. Since the interaction term in the Hubbard model acts only for
opposite spins it is clear that there is no renormalization  when the system is
completely polarized with one band fully occupied and the
other empty. The expectation value of the double occupancy
$\expval{n_{\uparrow}n_{\downarrow}}{}$ decreases with increasing field, which
further illustrates why the interaction term becomes less important for larger
fields. 

\begin{figure}[!htbp]
\centering
\includegraphics[width=0.45\textwidth]{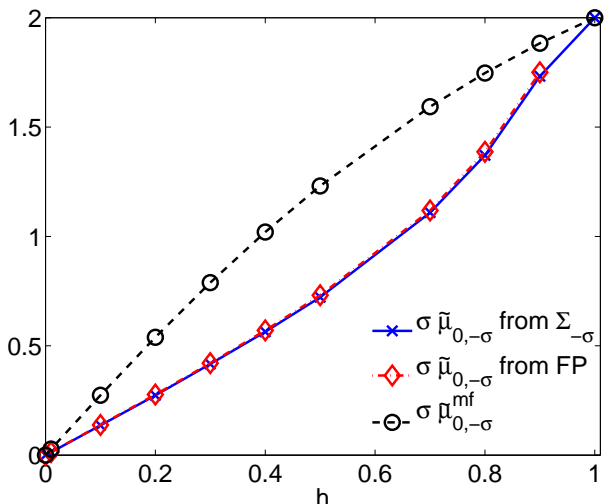}
\vspace*{-0.5cm}
\caption{(Color online) The renormalized chemical potential
  $\tilde\mu_{0,\sigma}(h)$ calculated from the impurity fixed point (FP) and
  directly from the self-energy for $U=2$ and various fields $h$. } 
\label{tmucompU2}
\end{figure}

We can also follow the field dependence of the renormalized chemical potential
$\tilde\mu_{0,\sigma}(h)$ as plotted in figure \ref{tmucompU2}. In the case of
particle hole symmetry we have
$\tilde\mu_{0,\uparrow}(h)=-\tilde\mu_{0,\downarrow}(h)$, and similar to the
case for the quasiparticle weight we have displayed the average of the up and
down spin values for each method. Also here the deviation within one method is
only due to small numerical inaccuracies and less then 2\%. 

\begin{figure}[!htbp]
\centering
\includegraphics[width=0.45\textwidth]{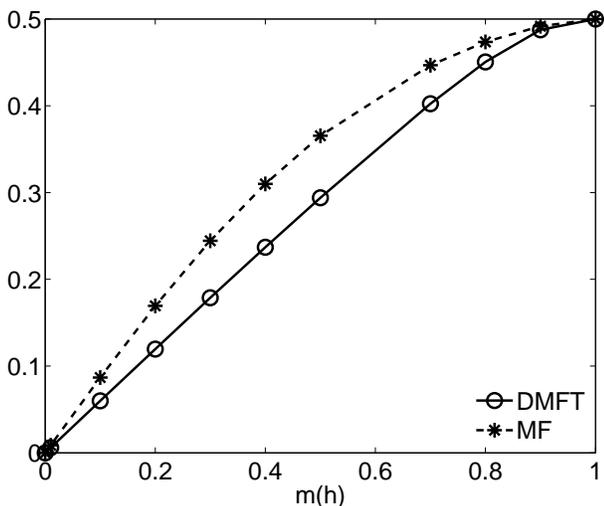}
\vspace*{-0.5cm}
\caption{The magnetization in the mean field approximation compared with the DMFT
  result for $U=2$ and for the full range of magnetic fields $h$. }
\label{magcompU2}
\end{figure}

\noindent
Again the agreement between the two methods of calculation 
is very good over the full range of magnetic fields. 
Mean field theory is valid for very weak interactions, so we compare 
our results for  $\tilde\mu_{0,\sigma}(h)$ for $U=2$ with the
mean field value  $\tilde\mu^{\rm mf}_{0,\sigma}=\mu+\sigma h
-Un^{\rm mf}_{-\sigma}$ in figure \ref{tmucompU2}.
The results coincide for $h=0$, when $\tilde\mu^{\rm
  mf}_{0,\sigma}=0$ and when the system becomes 
fully polarized at large field values, $\tilde\mu^{\rm
  mf}_{0,\sigma}=-\sigma(U+h)$, but in general  $\tilde\mu^{\rm
  mf}_{0,\sigma}>\tilde\mu_{0,\sigma}(h)$. We also compare the mean field (MF)
result for the field dependence of the magnetization $m(h)$  with the one
obtained in the DMFT calculation in figure \ref{magcompU2}. 

\noindent
The general behavior is similar, but the mean field theory without quantum
fluctuations overestimates the magnetization, as one would expect.

\subsection{Intermediate coupling regime}
\begin{figure}[!htbp]
\centering
\includegraphics[width=0.45\textwidth]{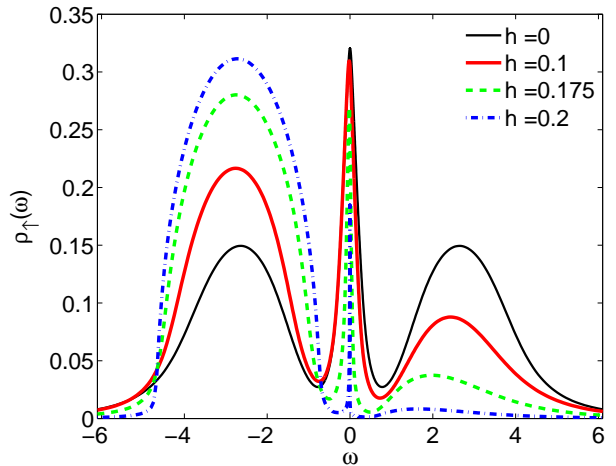}
\vspace*{-0.5cm}
\caption{(Color online) The local spectral density for the majority spin
  $\rho_\uparrow(\omega)$ for $U=5$ and various fields $h$. } 
\label{dosimchdep}
\end{figure}

In the next plot in figure \ref{dosimchdep}, where $U=5$, we show   typical
behavior of the local spectral density in the intermediate coupling  regime. 

\begin{figure}[!htbp]
\centering
\includegraphics[width=0.45\textwidth]{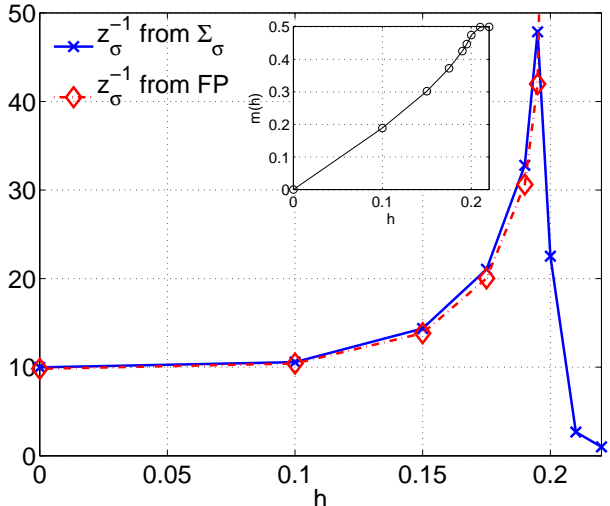}
\vspace*{-0.5cm}
\caption{(Color online) The inverse of the quasiparticle weight $z_\sigma(h)$
  calculated from the impurity fixed point (FP) and directly from the
  self-energy for $U=5$ and various fields $h$. The inset shows the
  magnetization $m(h)$.} 
\label{qpweightcompU5}
\end{figure}

\noindent
Similar to the weak coupling regime, we find a shift of spectral
weight towards lower energy for the majority spin. There is, however, a
difference in the way this happens due to the initial three peak structure,
namely the quasiparticle peak in the middle gets narrower for increasing field
and finally vanishes in the polarized phase. The quasiparticle weight,
which is shown in figure \ref{qpweightcompU5},  reflects
this behavior by decreasing to zero  with increasing field signaling
heavy quasiparticles. Here, as in the weak coupling case, we plot the average
of the spin up and down results for each method. The deviations can be larger
here, especially close to the metamagnetic transition.

\begin{figure}[!htbp]
\centering
\includegraphics[width=0.45\textwidth]{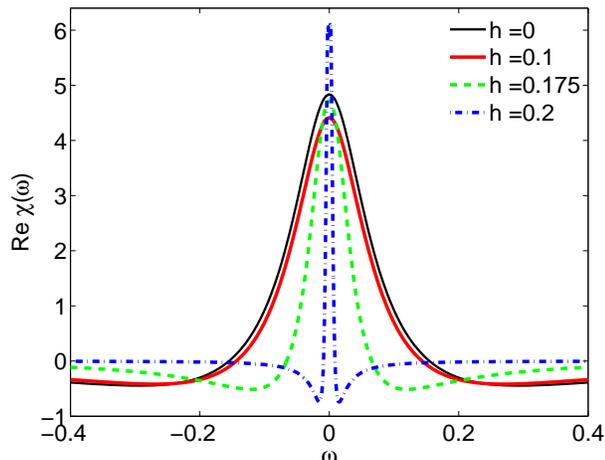}
\vspace*{-0.5cm}
\caption{(Color online) The real part of the local longitudinal dynamic spin
  susceptibility for $U=5$ and various fields $h$.} 
\label{suscwchdep}
\end{figure}

\noindent
When the material is polarized, $z_\sigma(h)$ reverts to 1, which corresponds
to the band insulator as before. This approach to 
the fully polarized localized state in high fields contrasts with that found 
in the weak coupling regime. It gives rise to metamagnetic behavior in this 
parameter regime. To  illustrate further the different response to a magnetic
field, the real  part of the local longitudinal dynamic spin susceptibility
$\chi_l(\omega,h)$ as a function  of $\omega$ is shown for various values of
$h$ in figure \ref{suscwchdep}. 

\noindent
It can be seen that the local susceptibility $\chi^{\mathrm{loc}}(h)={\rm
Re}\,\chi_l(0,h)$ in this regime increases with $h$ so that
$\partial\chi^{\mathrm{loc}}(h)/\partial h >0$. This can also be seen in the
curvature of the magnetization shown in the inset of figure
\ref{qpweightcompU5}. This is behavior characteristic of a  metamagnetic
transition and related to the magnetic field induced metal-insulator
transition. Laloux et al. \cite{LGK94} find metamagnetic behavior in a
similar parameter regime. There a comparison is made with results from  the
Gutzwiller approximation, which gives such a behavior already for smaller
values of the interaction, and we refer to their paper for details.

\begin{figure}[!htbp]
\centering
\includegraphics[width=0.45\textwidth]{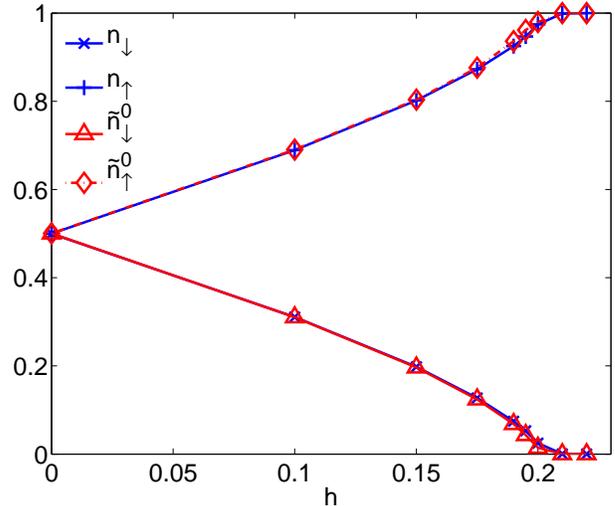}
\vspace*{-0.5cm}
\caption{(Color online) The comparison of the spin dependent occupation numbers $\tilde
  n_\sigma^0$ and $n_\sigma$ corresponding to Luttinger's theorem in a magnetic
  field for $U=5$ and the range of fields $h$.  }
\label{luttU5}
\end{figure}

We can also check the Luttinger theorem in a magnetic field (\ref{maglat})
by comparing the  values of $\tilde n_\sigma^0$,  deduced from integrating the
quasiparticle density of states, with the value of  $n_\sigma$ calculated from
the direct NRG evaluation in the ground state. The results are shown in
figure \ref{luttU5}.  

\noindent
It can be seen that there is excellent agreement between the results of these
two different calculations,  $\tilde n_\sigma^0=n_\sigma$, so that 
 Luttinger's theorem is satisfied for all values of the magnetic field in this
intermediate coupling regime.

\begin{figure}[!htbp]
\centering
\includegraphics[width=0.45\textwidth]{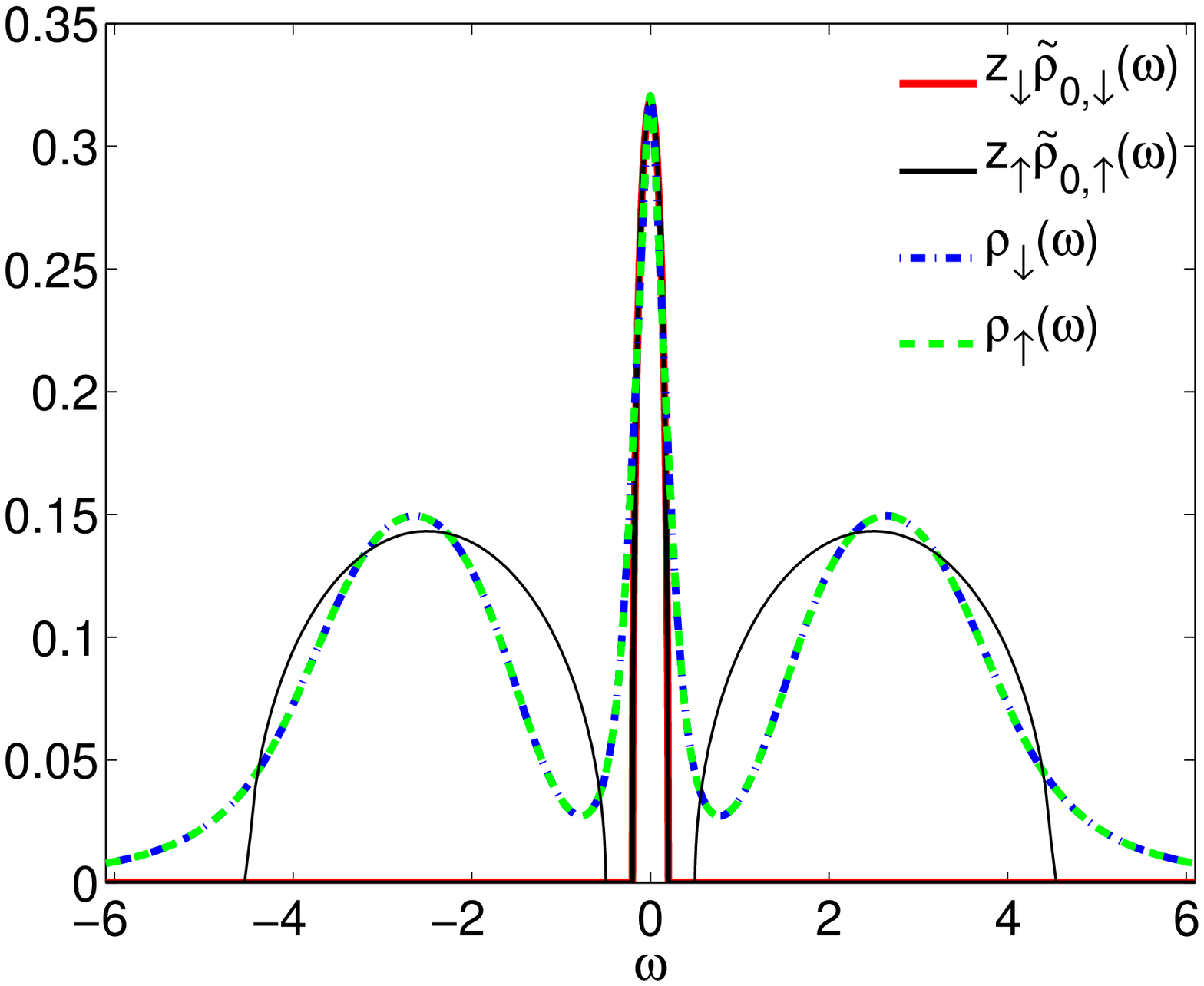}
\includegraphics[width=0.45\textwidth]{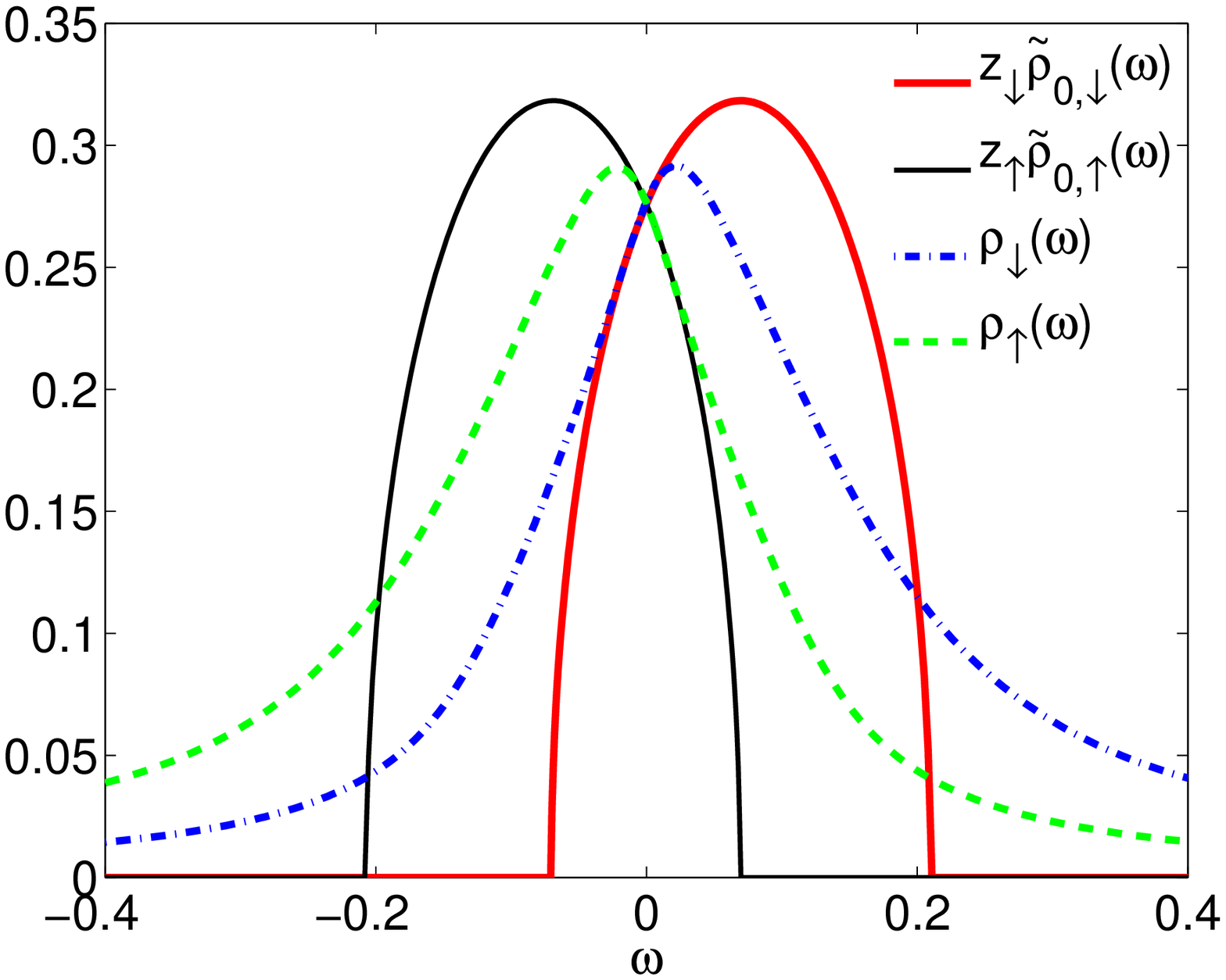}
\vspace*{-0.5cm}
\caption{(Color online) The free quasiparticle density of states in comparison
  with interacting local spectral density for $U=5$ and $h=0$ (upper
  panel). We have also plotted a thin black line for $\rho_{\rm
  mf}(\omega)=[D(\omega+U/2)+D(\omega-U/2)]/2$ which describes the
  non-magnetic mean field solution and weighted with $1-z_{\sigma}$. In the
  lower panel we have a similar comparison for $U=5$ and $h=0.15$
  concentrating on low energies.}    
\label{qpdosh_0.0}
\end{figure}

%\subsection{Quasiparticle dynamics}
Having deduced the renormalized parameters of the quasiparticles from the NRG
results, we are now in a position to test
how well we can describe the low energy dynamics of this model in a magnetic
field in terms of a renormalized perturbation theory. 
%\subsubsection{Free quasiparticle spectral density}
It is of interest first of all to see how  the free quasiparticle density of 
states $\tilde\rho_{0,\sigma}(\omega)$ multiplied by $z_\sigma(h)$ compares with
the full spectral density $\rho_\sigma(\omega)$. In figure \ref{qpdosh_0.0} (upper
panel) we make a comparison  in the zero magnetic field case. 

\noindent
We see that the quasiparticle band gives a good representation of the low
energy peak in $\rho_\sigma(\omega)$ and, as expected, does not reproduce the
high energy features. These, however, to a fair approximation can be described
by the mean field solution $\rho_{\rm mf}(\omega)$ weighted with a factor
$1-z_{\sigma}$ as can be seen in figure \ref{qpdosh_0.0} (upper panel).
A case with a finite magnetic field $h=0.15$, where the peaks in the density
of states of the two spin species are shifted due to the induced polarization 
relative to the Fermi level, is shown in figure \ref{qpdosh_0.0} (lower panel). 
%\begin{figure}[!htbp]
%\centering
%\includegraphics[width=0.45\textwidth]{figures_wfdm/qpDOScomp_U5_h0.15.eps}
%\vspace*{-0.5cm}
%\caption{The free quasiparticle density of states in comparison with interacting the
%  local spectral density   } 
%\label{qpdosh_0.15}
%\end{figure}
The figure focuses on the region at the Fermi level and one can see that the
free quasiparticle density of states describes well the form of
$\rho_\sigma(\omega)$ in the immediate vicinity of the Fermi level. It is to
be expected that the frequency range for this agreement can be extended if
self-energy corrections are included in the quasiparticle density of states
using the renormalized perturbation theory as shown in the impurity case
\cite{BHO07}.

%In the fully polarized case with $h=0.22$ there is complete agreement between the
%quasiparticle density of states and  $\rho_\sigma(\omega)$ for both spin types
%as can be seen in the results shown in \ref{qpdosh_0.22}, where
%$z_\uparrow=z_\downarrow=1$. Note that the parameters $\tilde\mu_{0,\sigma}$
%and $z_{\sigma}$ are purely derived from the NRG self-energy in this case.
%
%\begin{figure}[!htbp]
%\centering
%\includegraphics[width=0.45\textwidth]{figures_wfdm/qpDOScomp_U5_h0.22.eps}
%\vspace*{-0.5cm}
%\caption{The free quasiparticle density of states in comparison with interacting local
%  spectral density for $U=5$ and $h=0.22$. } 
%\label{qpdosh_0.22}
%\end{figure}
%\subsection{Dynamic susceptibilities at Half-filling}

\begin{figure}[!htbp]
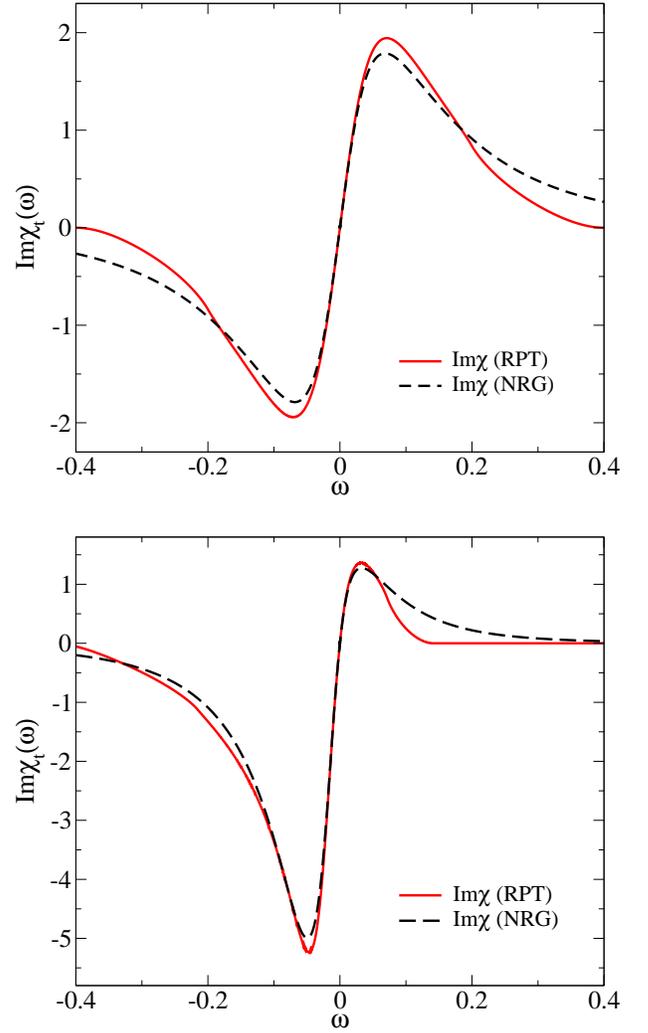

%\centering
\begin{center}
\includegraphics[width=0.45\textwidth]{figures_alex/latchit_U5.0_h_0.0.eps}\\[0.5cm]
%\vspace{0.5cm}
\includegraphics[width=0.45\textwidth]{figures_alex/latchit_U5.0_h_0.15.eps}
\end{center}
\vspace*{-0.5cm}
\caption{(Color online) A comparison of the imaginary parts of the  transverse
  dynamic spin susceptibility  for $U=5$ and $h=0.0$ (upper panel) and
  $h=0.15$ (lower panel) calculated using renormalized perturbation theory
  (RPT, full line)  and from a direct NRG calculation (dashed line). } 
\label{itchi0.0}
\end{figure}

We now compare the NRG results for the longitudinal and transverse local
dynamic spin susceptibilities for the same value $U=5$ and a similar range of
magnetic field values with those based on the RPT formulae (\ref{lrrpt})
and (\ref{trrpt}). In figure \ref{itchi0.0} (upper panel) we show the imaginary part of the
transverse spin susceptibility calculated with the two different methods.

\noindent
It can be seen that RPT formula gives the overall form of the NRG results,
and precisely fits  the gradient of the NRG curve at $\omega=0$.
Some of the relatively small differences between the results might be attributed to the
broadening factor used in the NRG results which gives a slower fall off with
$\omega$ in the higher frequency range, and a slightly reduced peak.
We get similar good agreement between the two sets of results for the same
quantity for the case with a magnetic field $h=0.15$, shown in figure
 \ref{itchi0.0} (lower panel).
%\ref{itchi0.15}. 

%\begin{figure}[!htbp]
%\centering
%\includegraphics[width=0.45\textwidth]{figures_alex/latchit_U5.0_h_0.15.eps}
%%\vspace*{-0.5cm}
%\caption{Plots of the imaginary part of the  transverse dynamic spin
%  susceptibility for $U=5$ and $h=0.15$. } 
%\label{itchi0.15}
%\end{figure}

\begin{figure}[!htbp]
\centering
\includegraphics[width=0.45\textwidth]{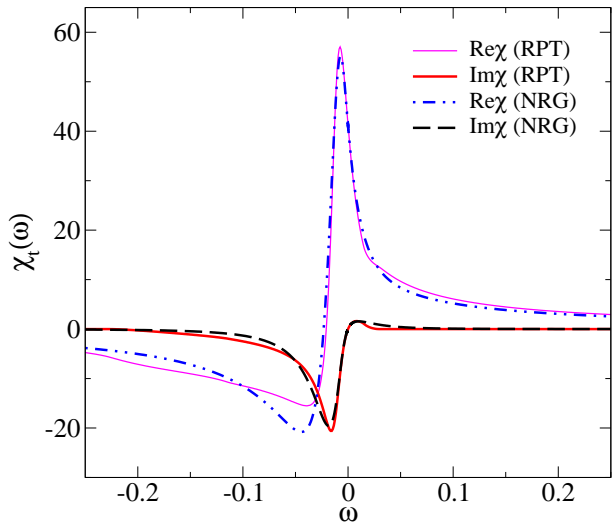}
\vspace*{-0.3cm}
\caption{(Color online) The real and imaginary parts of the transverse dynamic spin
  susceptibility  for $U=5$ and $h=0.19$. }  
\label{tchi0.19}
\end{figure}

In figure \ref{tchi0.19}, where we give both the real and imaginary parts of the transverse
susceptibility for $h=0.19$,  we see that this overall agreement is maintained
in the large field regime where we get metamagnetic behavior. The shapes of
the low energy peaks for 
both quantities are well reproduced by the RPT formulae. Note that the peak in 
the real part is not at $\omega=0$, so it is not fixed by the condition that
determines $\tilde U_t$, but nevertheless is in good agreement with the NRG
results. Due to their very small values it becomes difficult to calculate 
 $z_\sigma(h)$  as the system approaches localization for larger fields.
In this regime as $z_\sigma(h)\to 0$ the free  quasiparticle density
of states will converge to a delta-function. Self-energy corrections to the
free quasiparticle propagators, which were used in the calculation of
$\tilde\chi_{\sigma,\sigma'}(\omega)$, will become increasingly important as
this limit is approached. Once the system has undergone the localization
transition, and is completely polarized, however, we find 
%saw  in figure       %\ref{qpdosh_0.22}   
that the values $\tilde\mu_\sigma$ ($z_\sigma(h)=1$) deduced from the
self-energy give a quasiparticle density of states coinciding with the
DMFT-NRG result of an upper and lower semi-circular bands.

\begin{figure}[!htbp]
\centering
\includegraphics[width=0.45\textwidth]{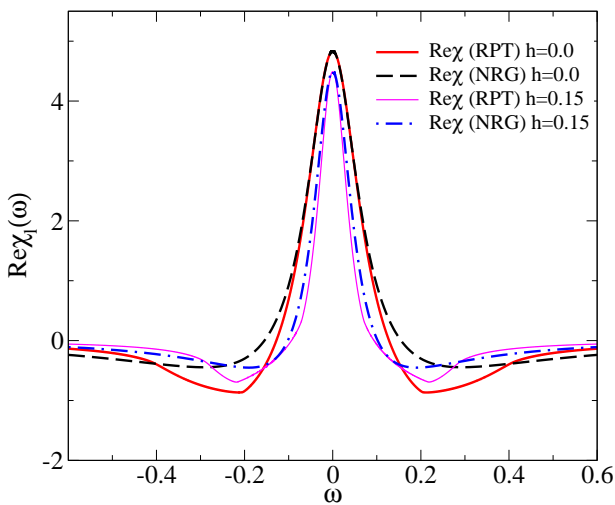}
\vspace*{-0.3cm}
\caption{(Color online) The real part of the  longitudinal dynamic spin susceptibility for
  $U=5$ and $h=0$ and $h=0.15$. }  
\label{rlchi}
\end{figure}

\begin{figure}[!htbp]
\centering
\includegraphics[width=0.45\textwidth]{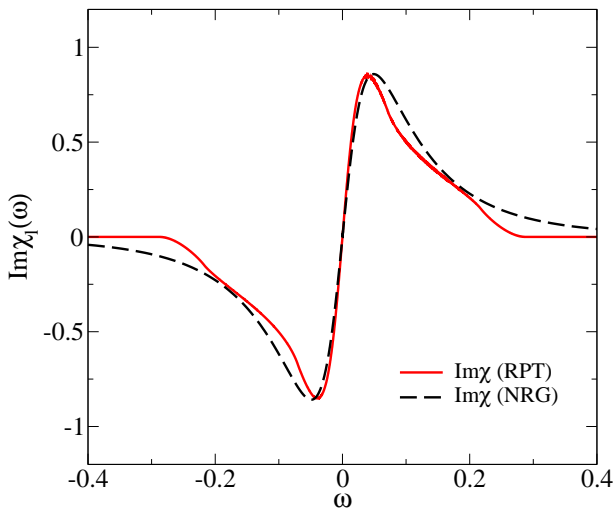}
\vspace*{-0.3cm}
\caption{(Color online) The imaginary part of the longitudinal dynamic spin susceptibility
  for $U=5$ and $h=0.15$. }  
\label{ilchi0.15}
\end{figure}

Results for the longitudinal susceptibility are shown in figures \ref{rlchi}
and  \ref{ilchi0.15}. In figure \ref{rlchi} we give the values for the real
part as a function of $\omega$ for $h=0$ and $h=0.15$.  

\noindent
Here the peak height, which is at $\omega=0$, is fixed by the condition which
determines $\tilde U_l$. The widths of the peaks in the two sets of NRG
results, however,  are given reasonably well by the RPT equations. The
imaginary part of the longitudinal susceptibility obtained by the two methods
is given in figure  \ref{ilchi0.15} for $h=0.15$. Again there is overall
agreement between the two sets of results. 
The slight undulations seen in the RPT results are due to the sharp cut off
in the band edges in the free quasiparticle density of states.
For larger values of $h$ the agreement with the NRG results is not as good
as that as for the transverse susceptibility, and the central peak in the
real part of the RPT results narrows more rapidly with $h$ than in those 
obtained from the direct NRG calculation.

%===========================================================================

\subsection{Strong coupling regime}
\begin{figure}[!htbp]
\centering
\includegraphics[width=0.45\textwidth]{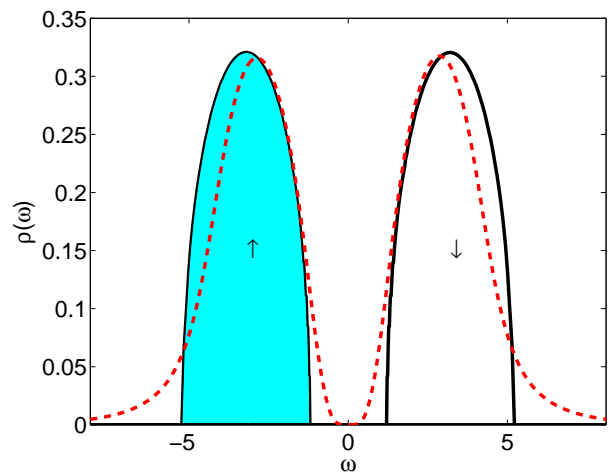}
\vspace*{-0.5cm}
\caption{(Color online) The total local spectral density $\rho(\omega)$ for $U=6$ for $h=0$
  (dashed line), Mott insulator,  and   $h=0.2$ (full line), fully polarized
  band insulator.  }
\label{dosmi}
\end{figure}

Finally we consider the strong coupling regime with $U>U_c$,
where for $h=0$ the spectral density has a Mott-Hubbard gap
so that for half-filling the system is an insulator. 
The electrons will be localized with free magnetic moments coupled by
an effective antiferromagnetic exchange $J\sim t^2/U$. In fields
such that $h>J$, the system polarizes completely as can be seen in figure
\ref{dosmi} where we show the total density of states
$\rho(\omega)=\rho_{\uparrow}(\omega) +\rho_{\downarrow}(\omega)$ for $h=0$
and $h=0.2$. 

\noindent
For smaller fields, such that $h<J$, we do not find a convergent
solution to  the DMFT equations, and the iterations oscillate between local
states which are either completely full or empty. We interpret this  as due to
the tendency to antiferromagnetic order which in a weak field, due to the
absence of anisotropy,  will be almost perpendicular to the applied field in
the x-y plane with a slight canting of the spins in the z-direction (spin
flopped phase). In this calculation no allowance has been made for this type of
ordering, but this state can be well described using an effective Heisenberg
model for the localized moments.

\section{Results away from Half filling}
After the extensive discussion of the behavior of the Hubbard model in a magnetic
field at half filling, we want to compare these results with the situation
where the system is doped with holes. As is well known, doping retains the
metallic character of the system and one does not find a paramagnetic
metal-insulator transition anymore. Thus we do not find distinct regimes
(a)-(c) any longer. To illustrate the characteristics of the magnetic response
for the doped system we focus on two cases, one at quarter filling and one very
close to half filling.

\subsection{Quarter Filled Case}

\begin{figure}[!htbp]
\centering
\includegraphics[width=0.45\textwidth]{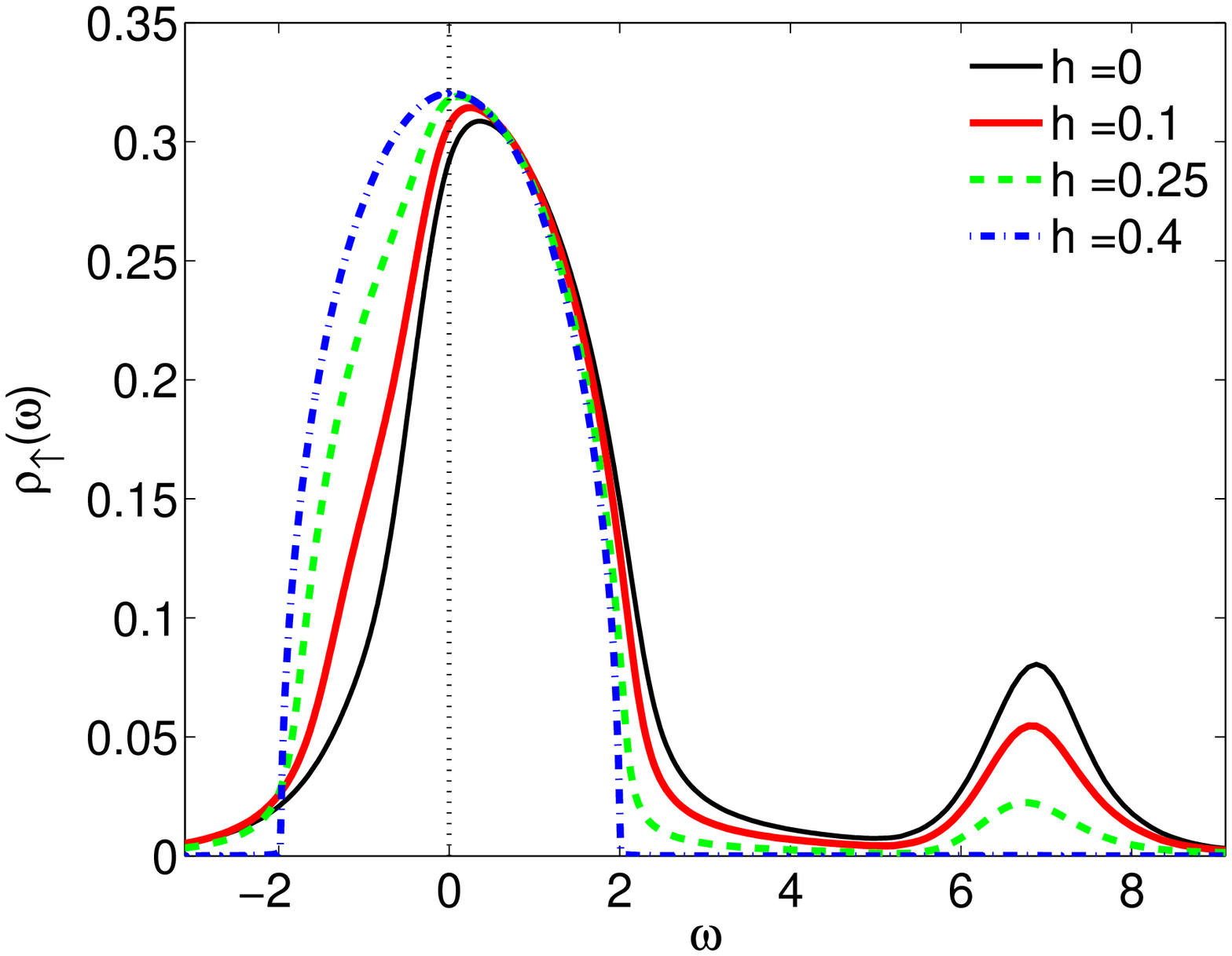}
\includegraphics[width=0.45\textwidth]{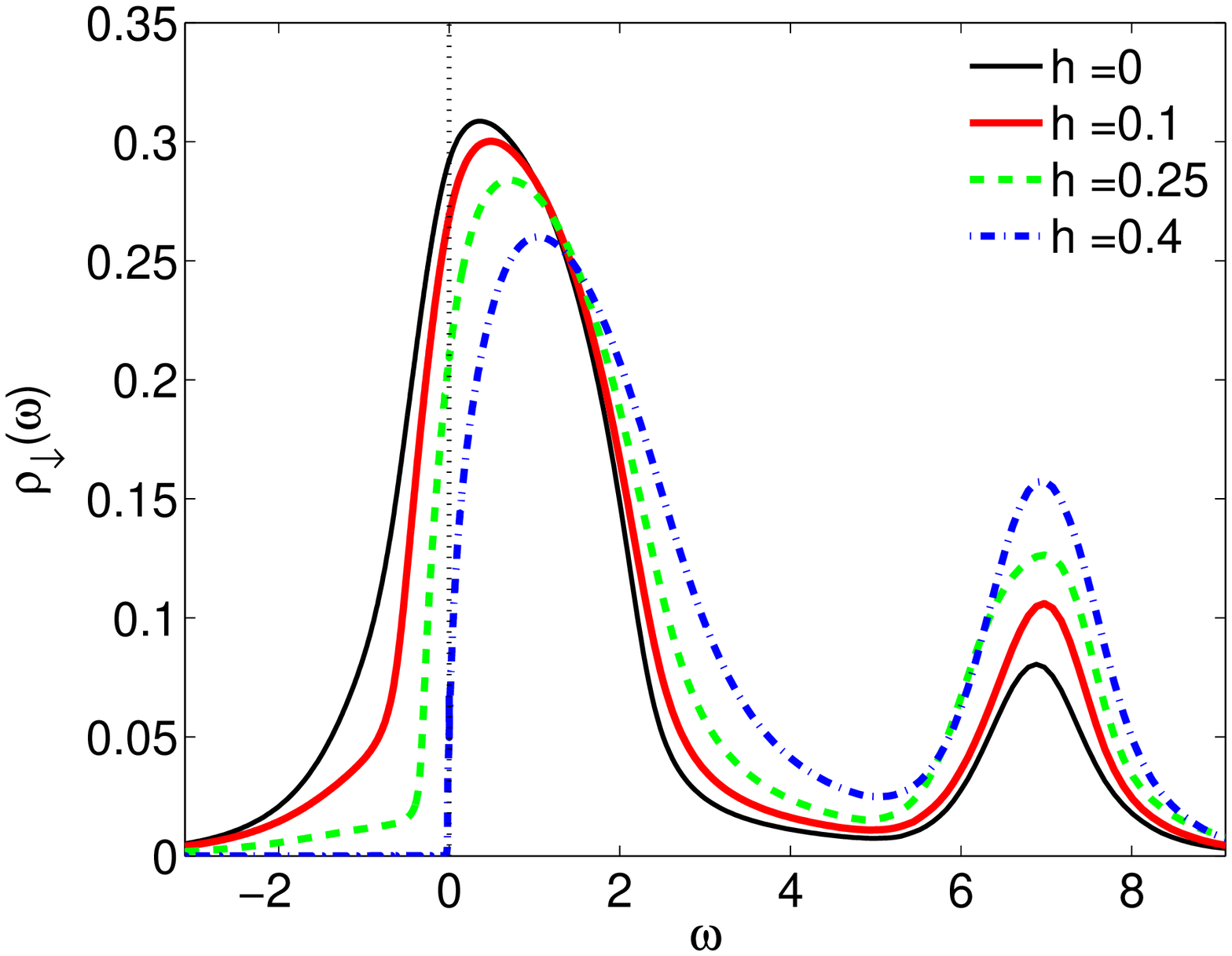}
\vspace*{-0.5cm}
\caption{(Color online) The local spectral density for the majority spin
  $\rho_{\uparrow}(\omega)$ (upper panel) and for the minority spin
  $\rho_{\downarrow}(\omega)$ (lower panel) for $U=5$, $x=0.5$ and various
  fields $h$. } 
\label{dosdopU5u}
\end{figure}

First we  compare the results in the intermediate coupling regime with $U=5$
at half-filling with those at quarter filling, $x=0.5$. In the latter case the
Fermi level falls in the lower Hubbard peak in the spectral density. To see
how the band changes with increasing magnetic field we plot the density of
states for both spin types, for the majority spin electrons  
in figure  \ref{dosdopU5u} (upper panel) and for the minority spin electrons in 
figure  \ref{dosdopU5u} (lower panel), for various values of the magnetic field. 

%\begin{figure}[!htbp]
%\centering
%\includegraphics[width=0.45\textwidth]{figures_wfdm/upDOS_x0.5diffh_U5.eps}
%\vspace*{-0.5cm}
%\caption{The local spectral density for the minority spin $\rho_{\downarrow}(\omega)$
%  for $U=5$, $x=0.5$ and various fields $h$. }
%\label{dosdopU5d}
%\end{figure}
\noindent
In the majority spin case the lower peak gains spectral weight on the low
energy side and the weight in the upper peaks decreases with increase of the
field. For the fully polarized case ($h>0.4$) the Fermi level, which is
indicated by a dotted line, lies in the middle of the lower band, which has the
non-interacting semi-circular shape. The opposite 
features can be seen in the minority spin case, with the spectral weight in
the lower peak below the Fermi level decreasing and the weight in the upper
peak increasing. Thus the increase of spectral weight below the Fermi level 
for the majority spin electrons, and the decrease for the minority 
spin electrons, can be seen to be due to a
change of band shape rather than a simple relative shift of the two bands,
which would be the case in  mean field theory.
In the fully polarized state there are no minority states
below the Fermi level and the upper peak in the majority state density of
states has disappeared. 

\begin{figure}[!htbp]
\centering
\includegraphics[width=0.45\textwidth]{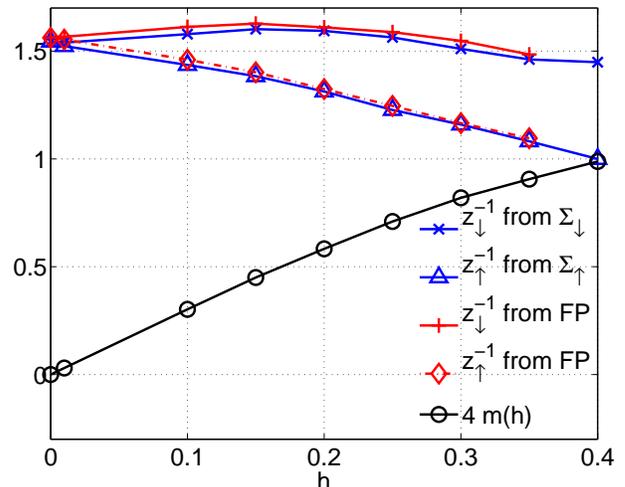}
\vspace*{-0.5cm}
\caption{(Color online) The inverse of the quasiparticle weight $z_\sigma(h)$ calculated from
  the impurity fixed point (FP) and directly from the self-energy for $U=5$, $x=0.5$ and
  various fields $h$. The magnetization $m(h)$ is also displayed.   }
\label{qpweightcompU5x.5}
\end{figure}

The corresponding values for the inverse of the quasiparticle weight
$1/z_\sigma(h)$ are shown in figure \ref{qpweightcompU5x.5} for a range of 
fields. 

\noindent
As noted in the impurity case \cite{BH07apre}, the quasiparticle weights differ for
the two spin types with $z_{\uparrow}(h)>z_{\downarrow}(h)$. The values of
$z_{\sigma}(h)$ have been 
calculated, as described earlier, both from the energy levels (FP) and from 
a numerical derivative NRG derived self-energy. There is reasonable
agreement between the two sets of results, and the small differences can
be attributed to the uncertainty in the numerical derivative of the NRG
self-energy. As observed in the case for the non-symmetric Anderson model
\cite{BH07apre}, there is an initial decrease of   $z_{\downarrow}(h)$ with
increase of $h$, whereas $z_{\uparrow}(h)$ increases monotonically.
This implies that the effective mass of the majority spin electrons decreases
to its bare value, whilst the effective mass of the minority spin electrons
does not decrease much. The reason for that is that in the polarized system the up
electrons cannot interact through the Hubbard interaction term, whereas a down
spin electron can interact with all the up spin electrons leading to an
enhanced mass.
The field dependence of the magnetization is also shown in figure
\ref{qpweightcompU5x.5}, and is similar to the half-filled case with a weak
interaction ($U=2$).  
We have calculated, but do not show, the corresponding occupation values for
$\tilde n_\sigma^0$ which again agree well with the values of  $\tilde n_\sigma$, confirming
Luttinger's theorem in a magnetic field. 

\begin{figure}[!htbp]
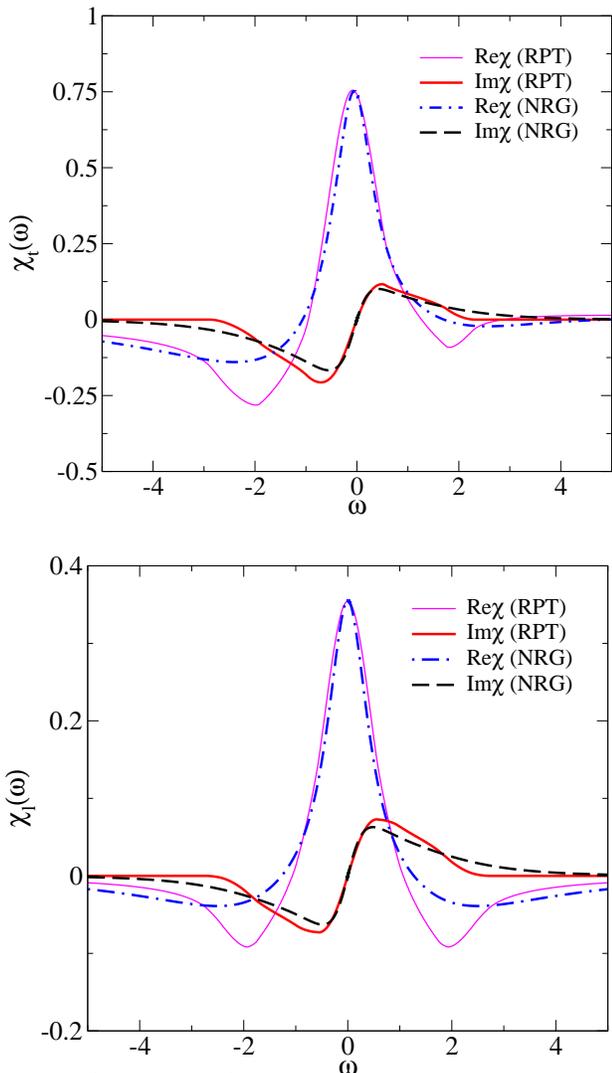

\centering
\includegraphics[width=0.45\textwidth]{figures_alex/latchit_U5.0_h_0.1_x_0.5.eps}\\[0.5cm]
\includegraphics[width=0.45\textwidth]{figures_alex/latchil_U5.0_h_0.1_x_0.5.eps}
\vspace*{-0.5cm}
\caption{(Color online) The real and imaginary parts of the  transverse
  dynamic spin susceptibility (upper panel) and of the longitudinal dynamic
  spin susceptibility (lower panel) for $U=5$, $x=0.5$  and $h=0.1$. } 
\label{tchix0.5}
\end{figure}

%\subsection{Dynamic susceptibilities at quarter filling}
We give two examples of  results for the susceptibilities for this case. 
In figure  \ref{tchix0.5} (upper panel) we plot the real and imaginary parts
of the transverse susceptibility. 

\noindent
Despite the large value of $U$, we can see that the peak heights are very much
reduced compared with those seen in the half-filled case for $U=5$. The peak
widths are also an order of magnitude larger as can be seen from the 
$\omega$-scale. The RPT results reproduce well the overall features to be
seen in the NRG results, but we note some discrepancies in the shape of the
curve at larger frequencies, where the RPT shows more pronounced features. The
real and imaginary parts for the longitudinal susceptibility are shown in
figure  \ref{tchix0.5} (lower panel).  
%\begin{figure}[!htbp]
%\centering
%\includegraphics[width=0.45\textwidth]{figures_alex/latchil_U5.0_h_0.1_x_0.5.eps}
%%\vspace*{-0.5cm}
%\caption{The real and imaginary parts of the longitudinal dynamic spin
%  susceptibility for $U=5$, $x=0.5$  and $h=0.1$. }  
%\label{lchix_0.5}
%\end{figure}
Again all the low energy features are reproduced in the RPT results and
differences are mainly seen for tails at larger energies. In this
regime, apart from the overall factor of 2, there is less difference between
the transverse and longitudinal susceptibilities than  at half-filling.

Our conclusion from these results, and from calculations with other values of
$U$ at quarter filling, is that when there is significant doping, the behavior in the field
corresponds to a weakly correlated Fermi liquid, very similar to that at half-filling
in the weak interaction regime. The only remarkable difference in the presence
of a magnetic field is the spin dependence of the effective masses as shown in
figure \ref{qpweightcompU5x.5}.

\subsection{Near half filling}

\begin{figure}[!htbp]
\centering
\includegraphics[width=0.45\textwidth]{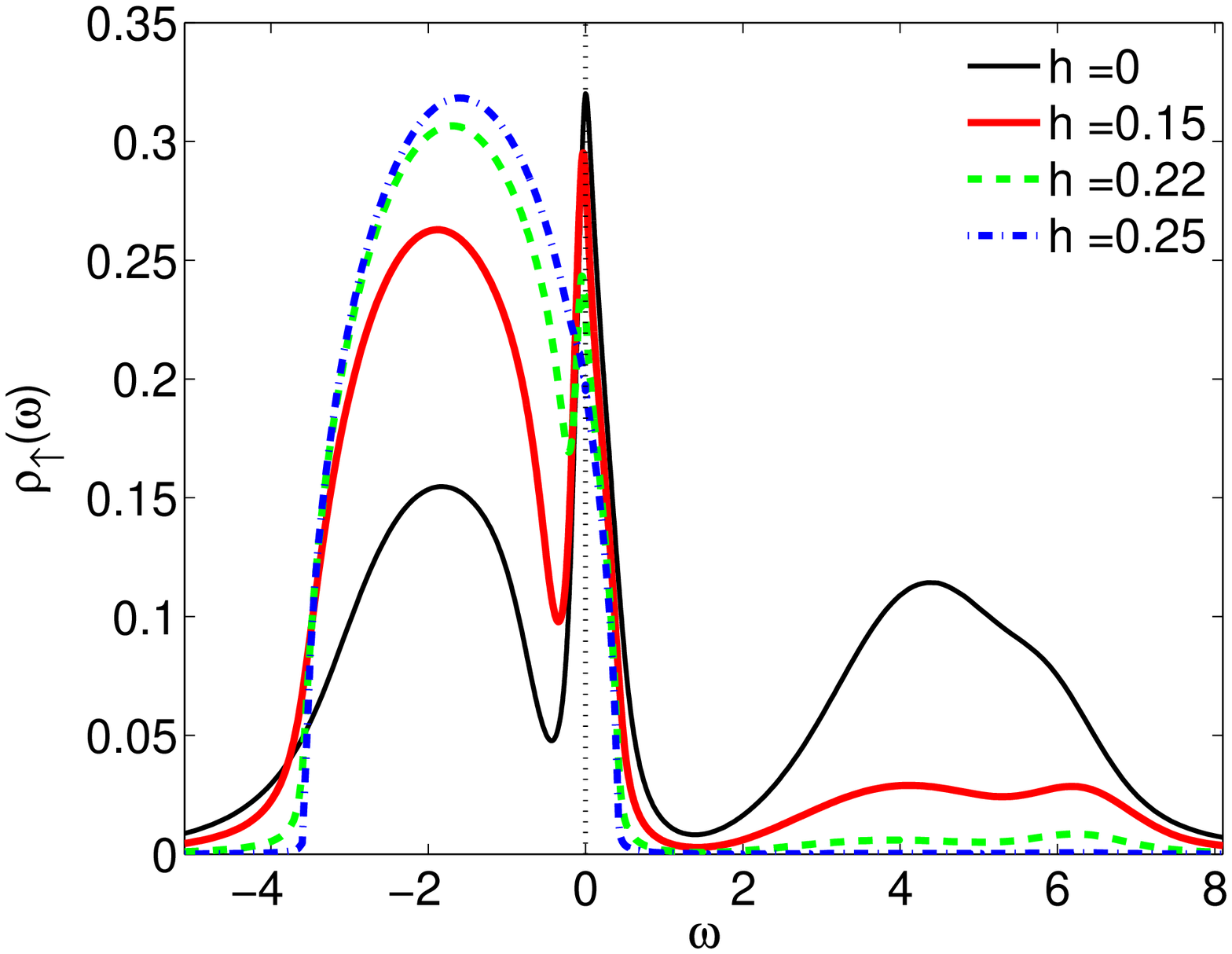}
\includegraphics[width=0.45\textwidth]{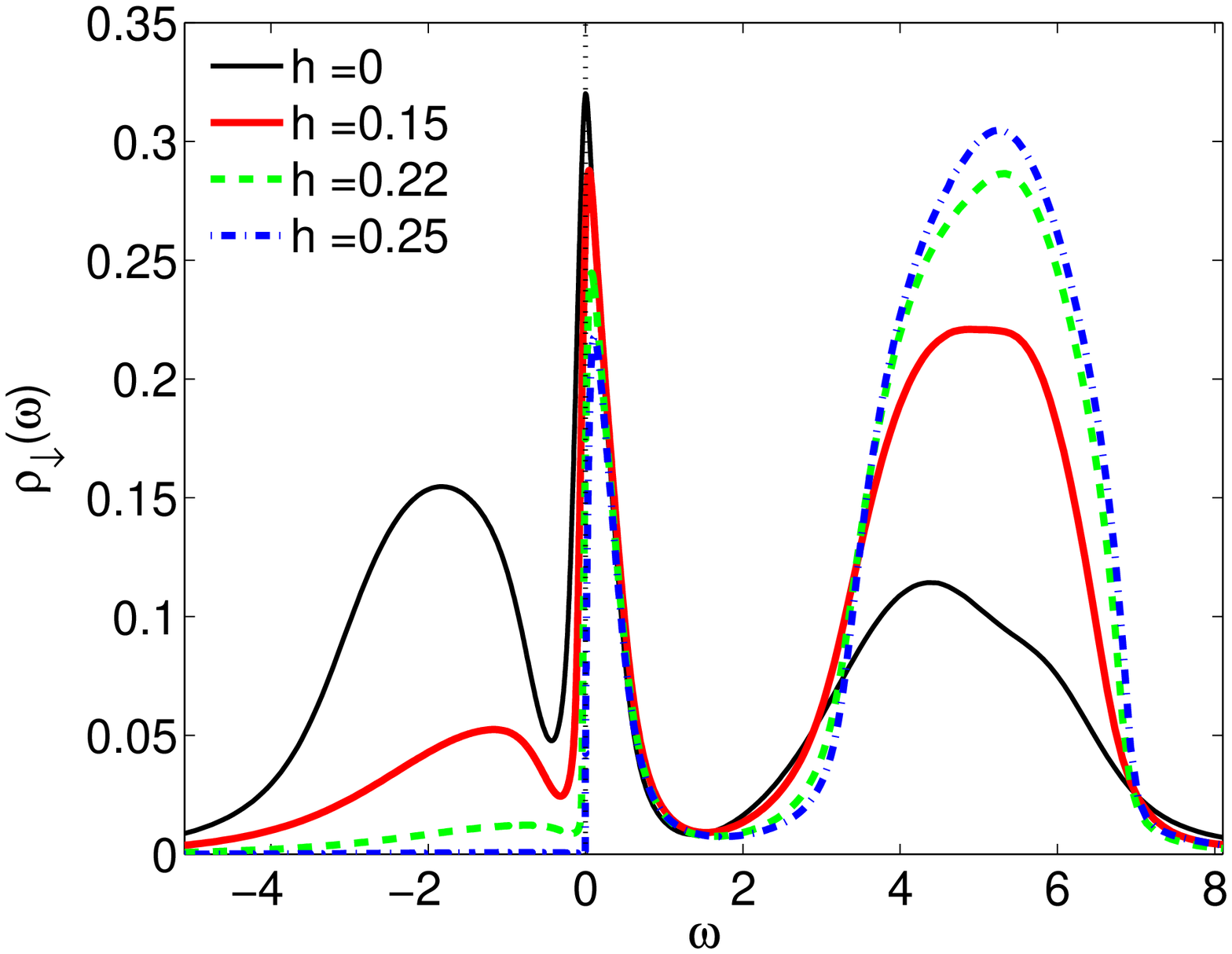}
\vspace*{-0.5cm}
\caption{(Color online) The local spectral density for the majority spin
  $\rho_{\uparrow}(\omega)$ (upper panel) and for the minority spin
  $\rho_{\downarrow}(\omega)$ (lower panel) for $U=6$, $x=0.95$ and various
  fields $h$. } 
\label{dosdopU6up}
\end{figure}

Very close to half-filling and for large values of $U$ we have a qualitatively
different parameter regime. Here the system is metallic but we can expect strong correlation
effects when $U$ is of the order or greater than $U_c$, due to the much 
reduced phase space for quasiparticle scattering. We look at the case 
with 5\% hole doping from half-filling and a value $U=6$, which is just
greater than the critical value for the metal-insulator transition. We show
the spectral density of states for the majority spin state in figure
\ref{dosdopU6up} (upper panel) and for the minority spin state 
in figure \ref{dosdopU6up} (lower panel) for various values of the magnetic
field.

\noindent
%\begin{figure}[!htbp]
%\centering
%\includegraphics[width=0.45\textwidth]{figures_wfdm/upDOS_x0.95diffh_U6.eps}
%\vspace*{-0.5cm}
%\caption{The local spectral density for the minority spin $\rho_{\downarrow}(\omega)$
%  for $U=6$, $x=0.95$ and various fields $h$. }
%\label{dosdopU6down}
%\end{figure}
There is a clear sharp quasiparticle peak for $h=0$ at the Fermi level (marked
by a dotted line) at the top of the lower Hubbard band. As in the quarter
filling case with $U=5$, we 
see a similar transfer of spectral weight with increasing field to below the
Fermi level for the majority  spin case, and above the Fermi level for the
minority spins. For large fields when the system is completely polarized the
Fermi level lies close to the top of the lower band in the majority spin
spectrum. One can see in the lower panel that there is still a sharp narrow
peak in the spectral density of the minority spin states above the Fermi
level, though the spectrum for the majority states 
below the Fermi level is that of the non-interacting system.  A spin up
electron added above the Fermi level feels no interaction as the system is
completely spin up polarized so these electrons see the non-interacting
density of states. On the other hand a spin down electron above the Fermi level
interacts strongly with the sea of up spin electrons. The self-energy due to
scattering with particle-hole pairs in the sea creates a distinct resonance in
the down spin density of states just above the Fermi level. Just such a
resonance was predicted by Hertz and Edwards \cite{HE72} for a Hubbard model
in a strong ferromagnetic (fully polarized) state. 

\begin{figure}[!htbp]
\centering
\includegraphics[width=0.45\textwidth]{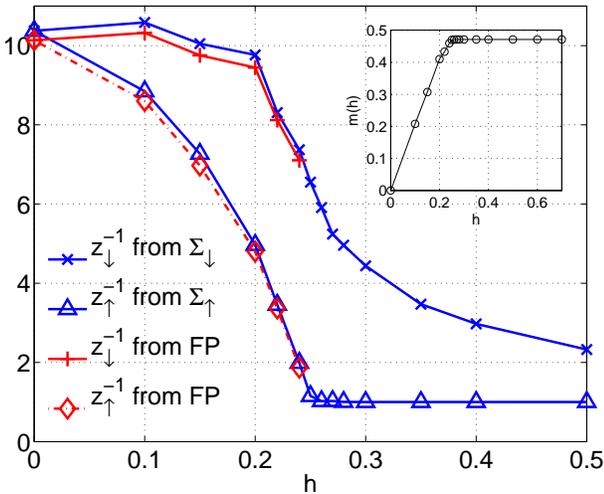}
\vspace*{-0.5cm}
\caption{(Color online) The inverse of the quasiparticle weight $z_\sigma(h)$ calculated
  from the impurity fixed point (FP) and directly from the self-energy for
  $U=6$, $x=0.95$ and various fields $h$. The inset shows the
  magnetization $m(h)$.   }
\label{qpweightcompU6x.5}
\end{figure}

The field dependence of the inverse of the quasiparticle weight is 
presented in figure \ref{qpweightcompU6x.5}.
 
\noindent
Again we find reasonable agreement between the two methods of calculation for these
quantities. The magnetization as a function of $h$ is shown as an inset in the
same figure. The behavior of  $z_\uparrow(h)$ and $z_\downarrow(h)$
as a function of $h$ contrasts sharply with the behavior found for the metallic
state at half-filling with $U=5$ shown in figure \ref{qpweightcompU5}. For
zero field the quasi-particle weight has a very similar value in both cases.
At half-filling the tendency of the magnetic field 
to induce localization resulted in values of $z_\uparrow(h)$ and
$z_\downarrow(h)$ ($z_\uparrow(h)=z_\downarrow(h)$) which decrease sharply
as a function of $h$. In the 5\% doped case with $U=6$, the system 
must remain metallic and the quasiparticles weights, $z_\uparrow(h)$ and
$z_\downarrow(h)$, both increase in large fields though their values differ
significantly. The quasiparticle weight for the minority spin electrons
decreases initially with increase of $h$, whereas that for the majority spins
 $z_\uparrow(h)$ increases monotonically and quite dramatically with $h$.
When the system becomes fully polarized ($h\simeq 0.26$) the up spin electrons become
essentially non-interacting, $z_\uparrow(h) = 1$, whereas there is a strong
interaction for a down spin electron and we find in this case
$z_\downarrow(h)\simeq 0.15$. The interpretation for this is as given in the
previous paragraph for the spectral densities. On further increasing the
magnetic field $z_\downarrow(h)$ also tends to one, but relatively slowly, as
can be seen in figure  \ref{qpweightcompU6x.5}. Note that the results in this
regime are based on the calculation from the self-energy, as the method based
on the fixed point analysis becomes difficult to apply in this regime.

Laloux et al. \cite{LGK94} compared the quasiparticle weight at half filling
for the infinite dimensional model with results from the Gutzwiller
approximation. The values for the infinite dimensional model were found to be
significantly smaller than the Gutzwiller predictions, the ratio is more
than a factor of $2$ for $U>4$ and zero field. Spa{\l}ek et
al. \cite{SG90,KSWA95} have made predictions based on the Gutzwiller approach
for situations away from half filling in finite field, $z_{\uparrow}\neq
z_{\downarrow}$. As in the study of Laloux et al. \cite{LGK94} our
results for $z_{\uparrow}$ and $z_{\downarrow}$ are significantly smaller
than the Gutzwiller predictions. 

\begin{figure}[!htbp]
\centering
\includegraphics[width=0.45\textwidth]{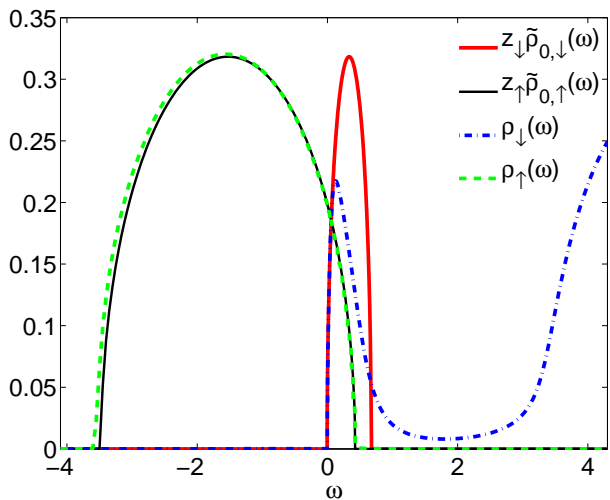}
\vspace*{-0.5cm}
\caption{(Color online) The free quasiparticle density of states in comparison
  with interacting local spectral density for $U=6$, $x=0.95$ and $h=0.26$. }  
\label{qpdosU6h_0.26}
\end{figure}

For the fully polarized  case ($h=0.26$) we show the comparison of the
weighted free quasiparticle density of states $z_{\sigma}\tilde
\rho_{0,\sigma}(\omega)$ with the full spectrum $\rho_{\sigma}(\omega)$ in
figure \ref{qpdosU6h_0.26}. 

\noindent
Note that the parameters $\tilde\mu_{0,\sigma}$ and $z_{\sigma}$ in $\tilde
\rho_{0,\sigma}(\omega)$ are purely derived from the NRG self-energy in this 
case. We can see that the different values for the field dependent
quasiparticle weight $z_{\sigma}(h)$ for up and down spin lead to remarkably
different quasiparticle band shapes. With $z_{\uparrow}\simeq 1$ the majority
spin quasiparticle  band is 
essentially that of the non-interacting density of states.
The very much smaller value  $z_{\downarrow}$ leads to a narrow quasiparticle
band above the Fermi level. The low energy flank of this quasiparticle band
describes well the narrow peak seen in the spectral density just above the
Fermi level. To describe these strong asymmetries in the spectral densities
near half filling, we need $z_{\uparrow}\gg z_{\downarrow}$, which contrasts
with the cases at half filling such as in figure \ref{qpdosh_0.0} where
always $z_{\uparrow}= z_{\downarrow}$. This suggests a discontinuous behavior of the
renormalization factors $z_{\sigma}$ on the approach to half filling.

\begin{figure}[!htbp]
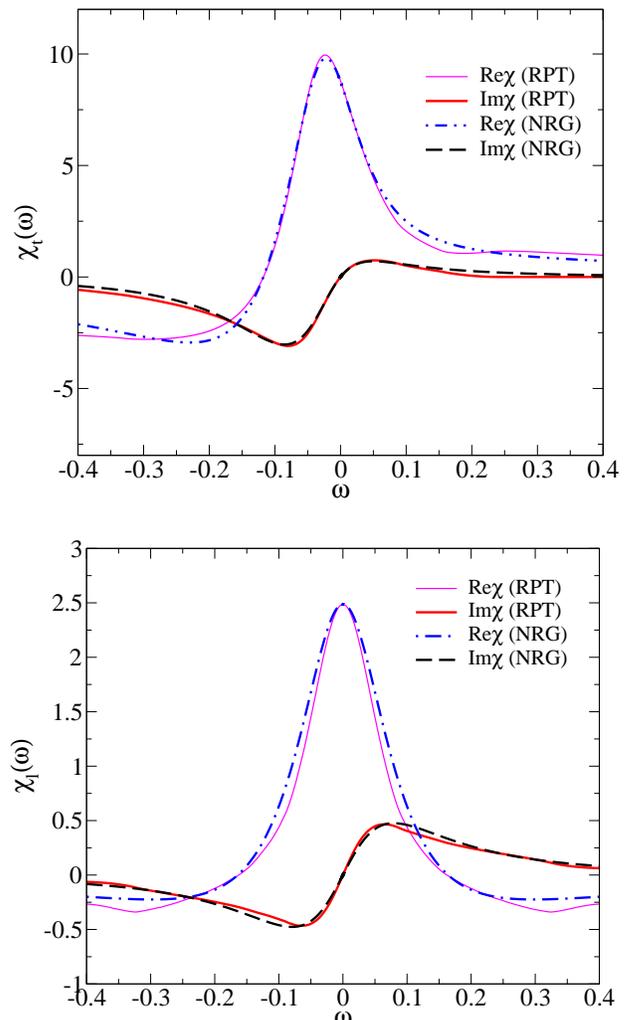

\centering
\includegraphics[width=0.45\textwidth]{figures_alex/latchit_U6.0_h_0.15_x_.95.eps}\\[0.5cm]
\includegraphics[width=0.45\textwidth]{figures_alex/latchil_U6.0_h_0.15_x_.95.eps}
\vspace*{-0.5cm}
\caption{(Color online) The real and imaginary parts of the transverse dynamic spin
  susceptibility (upper panel) and of the longitudinal dynamic spin
    susceptibility (lower panel) for $U=6$, $x=0.95$  and $h=0.15$. }  
\label{tchix0.95}
\end{figure}

Also for this case we show plots for the two susceptibilities for
a field of $h=0.15$. In figure \ref{tchix0.95} (upper panel) we give the real and
imaginary parts of the transverse susceptibility. 

\noindent
The low energy features are seen on an $\omega$-scale an order of magnitude
smaller than that for quarter filling due to the much stronger renormalization
effects in this regime. There is excellent agreement both with the peak positions and shapes
between the NRG and RPT results for both quantities. This is also seen to be
the case for the real and imaginary part of the longitudinal susceptibility
shown in figure \ref{tchix0.95} (lower panel), though the peak in the real part can be seen
to be marginally narrower in the RPT results. 

At the end of this section, we conclude that already a small doping of the
system is enough to maintain a metallic character even for very strong
interaction. Although the zero field spectra of the half filled case for $U=5$
and the small doping case with $U=6$ display very similar zero field behavior,
i.e. a  strongly renormalized quasiparticle band with similar $z_{\sigma}$, no
field induced localization transition occurs for finite doping and no
metamagnetic behavior is observed in the latter case.

%\begin{figure}[!htbp]
%\centering
%\includegraphics[width=0.45\textwidth]{figures_alex/latchil_U6.0_h_0.15_x_.95.eps}
%%\vspace*{-0.5cm}
%\caption{The real and imaginary parts of the longitudinal dynamic spin
%  susceptibility  for $U=6$, $x=0.95$  and $h=0.15$. }  
%\label{lchix0.95}
%\end{figure}

\section{Summary}

In this paper we have used the DMFT-NRG method to calculate the spectral
densities for one-particle and two-particle response functions for the 
infinite dimensional Hubbard model in a magnetic field, for the qualitatively
different filling regimes and interaction strengths. The results extend
earlier calculations of Laloux et al. \cite{LGK94} using the ED method,
which were restricted to the case of half-filling. Our results there
are on the whole consistent with this earlier work,  except in the insulating
regime for weak fields, where we could not find a convergent solution of the DMFT
equations. We attributed this to the fact that in this regime the magnetic
field is smaller than the exchange coupling between the localized spins so
that the ground state would be one in which the spins would have a  canted
antiferromagnetic ordering in the plane perpendicular to the field.

Well away from half filling we find a magnetic response similar to the weakly
correlated case even for large values of $U$. The large phase space for
quasiparticle scattering in this regime leads to modest renormalization
effects. Here, we find spin dependent quasiparticle weights,
$z_\uparrow(h)\ne z_\downarrow(h)$. This implies spin dependent as well as
field dependent effective masses, which  have been discussed earlier in work
by Spa{\l}ek et al. \cite{SG90,KSWA95} and Riseborough \cite{Ris06}. 
The calculations by Spa{\l}ek et al. were based on a Gutzwiller \cite{SG90} and a
mean-field slave boson approach \cite{KSWA95}. We can make a 
comparison of our results (section VI.B) near half filling, $x=0.95$, with theirs in the
later work \cite{KSWA95}. We find a qualitatively similar behavior with the
majority spin effective mass decreasing with $h$, but quantitatively there are
differences. The field dependence of the minority spin effective
mass $1/z_\downarrow(h)$ shows a
very slow increase initially in both sets of results, but the large field
behavior is quite different. As seen in figure \ref{qpweightcompU6x.5} we
find a significant decrease in $1/z_\downarrow(h)$ for large fields whereas
the corresponding quantity in figure 3 in reference \cite{KSWA95} increases.

The strong magnetic field dependence of the effective masses found in the
calculations by Riseborough is based on the assumption that the system is
close to a ferromagnetic transition (paramagnon theory).
However, DMFT calculations for the Hubbard model find that any ferromagnetism in
the Hubbard model only occurs in a very small region of the parameter space
near half-filling and for very large values of $U$. \cite{ZPB02}
Our results are well away from this regime and the large effective masses
obtained here can
be attributed to the tendency to localization rather than the tendency to
ferromagnetism.

Using the field dependent renormalized parameters $z_\sigma(h)$ and
$\tilde\mu_{0,\sigma}(h)$ in the RPT formulae for the dynamic local longitudinal
and transverse spin susceptibilities we found agreement with the overall
features to be seen in the DMFT-NRG results for these quantities. In the case
of the  transverse spin susceptibility excellent agreement was found
in all the metallic regimes and for all values of the magnetic field
considered, except in the high field regime at half filling  as
the localization point is approached, where consistent values of the
renormalized parameters are difficult to calculate. The comparison of the RPT
results with those from  NRG was also excellent for the longitudinal dynamic
susceptibility in the weaker field regime $h\le 0.15$ but less good for higher
fields, $h>0.15$.\par 

In all metallic parameter regimes a spin dependent Luttinger theorem in the
form $n_{\sigma}=\tilde n^0_{\sigma}$, the number of particles equals the
number of quasiparticles, was found to be satisfied for all strengths of the
magnetic field. In this form it even holds in the fully polarized insulating
state.
\par 

Phenomena like field and spin dependent effective masses and  metamagnetic
behavior have been observed experimentally in several heavy fermion compounds
\cite{AUAO93,GHTYF99,MCCSRC00,DBTWMB06}.
The Hubbard model, however, being a one band model is not an appropriate
starting point to make a quantitative comparison with the heavy fermion class
of materials. 
A periodic Anderson model with a two band structure and including the
degeneracy of the f electrons would be a better model to describe these
materials. Field dependent effects in this model have been studied by several
techniques, modified perturbation theory \cite{MN01}, exact diagonalization
\cite{SI96}, $1/N$ expansion \cite{Ono98} and variational approach \cite{EG97}.
The approach used here could be generalized to the periodic Anderson model,
but restricted to the non-degenerate case and $N=2$ as it is computationally 
too demanding in the NRG to deal with higher degeneracy.
The Hubbard model at half filling has been used as a lattice model to describe
the strongly renormalized Fermi liquid $^3$He \cite{Vol84,LGK94}. However, the
metamagnetic behavior predicted for relevant parameter regime is not seen
experimentally \cite{BFSWRPW98}. 
In section VI.B we found for small doping large effective masses, but no metamagnetic
behavior. This raises the possibility that the weakly doped Hubbard model
could serve as a basis for interpreting the experimental results for liquid
$^3$He.

\bigskip
\bigskip
\noindent{\bf Acknowledgment}\par
\bigskip
\noindent
We wish to thank N. Dupuis, D.M. Edwards, W. Koller, D. Meyer and A. Oguri for helpful
discussions and   W. Koller and D. Meyer for their contributions to the
development of the NRG programs. 
One of us (J.B.) thanks the Gottlieb Daimler and Karl Benz Foundation, the
German Academic Exchange Service (DAAD) and the EPSRC for financial support.

\bibliography{artikel,biblio1}

\begin{thebibliography}{36}
\expandafter\ifx\csname natexlab\endcsname\relax\def\natexlab#1{#1}\fi
\expandafter\ifx\csname bibnamefont\endcsname\relax
  \def\bibnamefont#1{#1}\fi
\expandafter\ifx\csname bibfnamefont\endcsname\relax
  \def\bibfnamefont#1{#1}\fi
\expandafter\ifx\csname citenamefont\endcsname\relax
  \def\citenamefont#1{#1}\fi
\expandafter\ifx\csname url\endcsname\relax
  \def\url#1{\texttt{#1}}\fi
\expandafter\ifx\csname urlprefix\endcsname\relax\def\urlprefix{URL }\fi
\providecommand{\bibinfo}[2]{#2}
\providecommand{\eprint}[2][]{\url{#2}}

\bibitem[{\citenamefont{Joss et~al.}(1987)\citenamefont{Joss, van Ruitenbeek,
  Crabtree, Tholence, van Deursen, and Fisk}}]{JRCTDF87}
\bibinfo{author}{\bibfnamefont{W.}~\bibnamefont{Joss}},
  \bibinfo{author}{\bibfnamefont{J.~M.} \bibnamefont{van Ruitenbeek}},
  \bibinfo{author}{\bibfnamefont{G.~W.} \bibnamefont{Crabtree}},
  \bibinfo{author}{\bibfnamefont{J.~L.} \bibnamefont{Tholence}},
  \bibinfo{author}{\bibfnamefont{A.~P.~J.} \bibnamefont{van Deursen}},
  \bibnamefont{and} \bibinfo{author}{\bibfnamefont{Z.}~\bibnamefont{Fisk}},
  \bibinfo{journal}{Phys. Rev. Lett.} \textbf{\bibinfo{volume}{59}},
  \bibinfo{pages}{1609} (\bibinfo{year}{1987}).

\bibitem[{\citenamefont{Goodrich et~al.}(1999)\citenamefont{Goodrich, Harrison,
  Teklu, Young, and Fisk}}]{GHTYF99}
\bibinfo{author}{\bibfnamefont{R.~G.} \bibnamefont{Goodrich}},
  \bibinfo{author}{\bibfnamefont{N.}~\bibnamefont{Harrison}},
  \bibinfo{author}{\bibfnamefont{A.}~\bibnamefont{Teklu}},
  \bibinfo{author}{\bibfnamefont{D.}~\bibnamefont{Young}}, \bibnamefont{and}
  \bibinfo{author}{\bibfnamefont{Z.}~\bibnamefont{Fisk}},
  \bibinfo{journal}{Phys. Rev. Lett.} \textbf{\bibinfo{volume}{82}},
  \bibinfo{pages}{3669} (\bibinfo{year}{1999}).

\bibitem[{\citenamefont{Aoki et~al.}(1993)\citenamefont{Aoki, Uji, Albessard,
  and Onuki}}]{AUAO93}
\bibinfo{author}{\bibfnamefont{H.}~\bibnamefont{Aoki}},
  \bibinfo{author}{\bibfnamefont{S.}~\bibnamefont{Uji}},
  \bibinfo{author}{\bibfnamefont{A.~K.} \bibnamefont{Albessard}},
  \bibnamefont{and} \bibinfo{author}{\bibfnamefont{Y.}~\bibnamefont{Onuki}},
  \bibinfo{journal}{Phys. Rev. Lett.} \textbf{\bibinfo{volume}{71}},
  \bibinfo{pages}{2110} (\bibinfo{year}{1993}).

\bibitem[{\citenamefont{Korbel et~al.}(1995)\citenamefont{Korbel, Spa{\l}ek,
  W\'ojcik, and Acquarone}}]{KSWA95}
\bibinfo{author}{\bibfnamefont{P.}~\bibnamefont{Korbel}},
  \bibinfo{author}{\bibfnamefont{J.}~\bibnamefont{Spa{\l}ek}},
  \bibinfo{author}{\bibfnamefont{W.}~\bibnamefont{W\'ojcik}}, \bibnamefont{and}
  \bibinfo{author}{\bibfnamefont{M.}~\bibnamefont{Acquarone}},
  \bibinfo{journal}{Phys. Rev. B} \textbf{\bibinfo{volume}{52}},
  \bibinfo{pages}{R2213} (\bibinfo{year}{1995}).

\bibitem[{\citenamefont{Manekar et~al.}(2000)\citenamefont{Manekar, Chaudhary,
  Chattopadhyay, Singh, Roy, and Chaddah}}]{MCCSRC00}
\bibinfo{author}{\bibfnamefont{M.}~\bibnamefont{Manekar}},
  \bibinfo{author}{\bibfnamefont{S.}~\bibnamefont{Chaudhary}},
  \bibinfo{author}{\bibfnamefont{M.~K.} \bibnamefont{Chattopadhyay}},
  \bibinfo{author}{\bibfnamefont{K.~J.} \bibnamefont{Singh}},
  \bibinfo{author}{\bibfnamefont{S.~B.} \bibnamefont{Roy}}, \bibnamefont{and}
  \bibinfo{author}{\bibfnamefont{P.}~\bibnamefont{Chaddah}},
  \bibinfo{journal}{J. Phys.: Cond. Mat.} \textbf{\bibinfo{volume}{12}},
  \bibinfo{pages}{9645} (\bibinfo{year}{2000}).

\bibitem[{\citenamefont{Vollhardt}(1984)}]{Vol84}
\bibinfo{author}{\bibfnamefont{D.}~\bibnamefont{Vollhardt}},
  \bibinfo{journal}{Rev. Mod. Phys.} \textbf{\bibinfo{volume}{56}},
  \bibinfo{pages}{99} (\bibinfo{year}{1984}).

\bibitem[{\citenamefont{Laloux et~al.}(1994)\citenamefont{Laloux, Georges, and
  Krauth}}]{LGK94}
\bibinfo{author}{\bibfnamefont{L.}~\bibnamefont{Laloux}},
  \bibinfo{author}{\bibfnamefont{A.}~\bibnamefont{Georges}}, \bibnamefont{and}
  \bibinfo{author}{\bibfnamefont{W.}~\bibnamefont{Krauth}},
  \bibinfo{journal}{Phys. Rev. B} \textbf{\bibinfo{volume}{50}},
  \bibinfo{pages}{3092} (\bibinfo{year}{1994}).

\bibitem[{\citenamefont{Janis and Czycholl}(2000)}]{JC00}
\bibinfo{author}{\bibfnamefont{V.}~\bibnamefont{Janis}} \bibnamefont{and}
  \bibinfo{author}{\bibfnamefont{G.}~\bibnamefont{Czycholl}},
  \bibinfo{journal}{Phys. Rev. B} \textbf{\bibinfo{volume}{61}},
  \bibinfo{pages}{9875} (\bibinfo{year}{2000}).

\bibitem[{\citenamefont{Kagawa et~al.}(2004)\citenamefont{Kagawa, Itou,
  Miyagawa, and Kanoda}}]{KIMK04}
\bibinfo{author}{\bibfnamefont{F.}~\bibnamefont{Kagawa}},
  \bibinfo{author}{\bibfnamefont{T.}~\bibnamefont{Itou}},
  \bibinfo{author}{\bibfnamefont{K.}~\bibnamefont{Miyagawa}}, \bibnamefont{and}
  \bibinfo{author}{\bibfnamefont{K.}~\bibnamefont{Kanoda}},
  \bibinfo{journal}{Phys. Rev. Lett.} \textbf{\bibinfo{volume}{93}},
  \bibinfo{pages}{127001} (\bibinfo{year}{2004}).

\bibitem[{\citenamefont{Hewson et~al.}(2006)\citenamefont{Hewson, Bauer, and
  Koller}}]{HBK06}
\bibinfo{author}{\bibfnamefont{A.~C.} \bibnamefont{Hewson}},
  \bibinfo{author}{\bibfnamefont{J.}~\bibnamefont{Bauer}}, \bibnamefont{and}
  \bibinfo{author}{\bibfnamefont{W.}~\bibnamefont{Koller}},
  \bibinfo{journal}{Phys. Rev. B} \textbf{\bibinfo{volume}{73}},
  \bibinfo{pages}{045117} (\bibinfo{year}{2006}).

\bibitem[{\citenamefont{Bauer and Hewson}(2007)}]{BH07apre}
\bibinfo{author}{\bibfnamefont{J.}~\bibnamefont{Bauer}} \bibnamefont{and}
  \bibinfo{author}{\bibfnamefont{A.~C.} \bibnamefont{Hewson}}
  (\bibinfo{year}{2007}), \bibinfo{note}{cond-mat/0705.3818}.

\bibitem[{\citenamefont{Hewson}(1993)}]{Hew93}
\bibinfo{author}{\bibfnamefont{A.~C.} \bibnamefont{Hewson}},
  \bibinfo{journal}{Phys. Rev. Lett.} \textbf{\bibinfo{volume}{70}},
  \bibinfo{pages}{4007} (\bibinfo{year}{1993}).

\bibitem[{\citenamefont{Hewson}(2001)}]{Hew01}
\bibinfo{author}{\bibfnamefont{A.~C.} \bibnamefont{Hewson}},
  \bibinfo{journal}{J. Phys.: Cond. Mat.} \textbf{\bibinfo{volume}{13}},
  \bibinfo{pages}{10011} (\bibinfo{year}{2001}).

\bibitem[{\citenamefont{Hewson}(2006)}]{Hew06}
\bibinfo{author}{\bibfnamefont{A.~C.} \bibnamefont{Hewson}},
  \bibinfo{journal}{J. Phys.: Cond. Mat.} \textbf{\bibinfo{volume}{18}},
  \bibinfo{pages}{1815} (\bibinfo{year}{2006}).

\bibitem[{\citenamefont{Metzner and Vollhardt}(1989)}]{MV89}
\bibinfo{author}{\bibfnamefont{W.}~\bibnamefont{Metzner}} \bibnamefont{and}
  \bibinfo{author}{\bibfnamefont{D.}~\bibnamefont{Vollhardt}},
  \bibinfo{journal}{Phys. Rev. Lett.} \textbf{\bibinfo{volume}{62}},
  \bibinfo{pages}{324} (\bibinfo{year}{1989}).

\bibitem[{\citenamefont{M{\"u}ller-Hartmann}(1989)}]{Mue89}
\bibinfo{author}{\bibfnamefont{E.}~\bibnamefont{M{\"u}ller-Hartmann}},
  \bibinfo{journal}{Z. Phys. B} \textbf{\bibinfo{volume}{74}},
  \bibinfo{pages}{507} (\bibinfo{year}{1989}).

\bibitem[{\citenamefont{Georges et~al.}(1996)\citenamefont{Georges, Kotliar,
  Krauth, and Rozenberg}}]{GKKR96}
\bibinfo{author}{\bibfnamefont{A.}~\bibnamefont{Georges}},
  \bibinfo{author}{\bibfnamefont{G.}~\bibnamefont{Kotliar}},
  \bibinfo{author}{\bibfnamefont{W.}~\bibnamefont{Krauth}}, \bibnamefont{and}
  \bibinfo{author}{\bibfnamefont{M.}~\bibnamefont{Rozenberg}},
  \bibinfo{journal}{Rev. Mod. Phys.} \textbf{\bibinfo{volume}{68}},
  \bibinfo{pages}{13} (\bibinfo{year}{1996}).

\bibitem[{\citenamefont{Luttinger}(1960)}]{Lut60}
\bibinfo{author}{\bibfnamefont{J.~M.} \bibnamefont{Luttinger}},
  \bibinfo{journal}{Phys. Rev.} \textbf{\bibinfo{volume}{119}},
  \bibinfo{pages}{1153} (\bibinfo{year}{1960}).

\bibitem[{\citenamefont{Krishna-murthy
  et~al.}(1980)\citenamefont{Krishna-murthy, Wilkins, and Wilson}}]{KWW80a}
\bibinfo{author}{\bibfnamefont{H.~R.} \bibnamefont{Krishna-murthy}},
  \bibinfo{author}{\bibfnamefont{J.~W.} \bibnamefont{Wilkins}},
  \bibnamefont{and} \bibinfo{author}{\bibfnamefont{K.~G.}
  \bibnamefont{Wilson}}, \bibinfo{journal}{Phys. Rev. B}
  \textbf{\bibinfo{volume}{21}}, \bibinfo{pages}{1003} (\bibinfo{year}{1980}).

\bibitem[{\citenamefont{Hewson et~al.}(2004)\citenamefont{Hewson, Oguri, and
  Meyer}}]{HOM04}
\bibinfo{author}{\bibfnamefont{A.~C.} \bibnamefont{Hewson}},
  \bibinfo{author}{\bibfnamefont{A.}~\bibnamefont{Oguri}}, \bibnamefont{and}
  \bibinfo{author}{\bibfnamefont{D.}~\bibnamefont{Meyer}},
  \bibinfo{journal}{Eur. Phys. J. B} \textbf{\bibinfo{volume}{40}},
  \bibinfo{pages}{177} (\bibinfo{year}{2004}).

\bibitem[{\citenamefont{Bulla et~al.}(1998)\citenamefont{Bulla, Hewson, and
  Pruschke}}]{BHP98}
\bibinfo{author}{\bibfnamefont{R.}~\bibnamefont{Bulla}},
  \bibinfo{author}{\bibfnamefont{A.~C.} \bibnamefont{Hewson}},
  \bibnamefont{and} \bibinfo{author}{\bibfnamefont{T.}~\bibnamefont{Pruschke}},
  \bibinfo{journal}{J. Phys.: Cond. Mat.} \textbf{\bibinfo{volume}{10}},
  \bibinfo{pages}{8365} (\bibinfo{year}{1998}).

\bibitem[{\citenamefont{Peters et~al.}(2006)\citenamefont{Peters, Pruschke, and
  Anders}}]{PPA06}
\bibinfo{author}{\bibfnamefont{R.}~\bibnamefont{Peters}},
  \bibinfo{author}{\bibfnamefont{T.}~\bibnamefont{Pruschke}}, \bibnamefont{and}
  \bibinfo{author}{\bibfnamefont{F.~B.} \bibnamefont{Anders}},
  \bibinfo{journal}{Phys. Rev. B} \textbf{\bibinfo{volume}{74}},
  \bibinfo{pages}{245114} (\bibinfo{year}{2006}).

\bibitem[{\citenamefont{Weichselbaum and {von Delft}}(2006)}]{WD06pre}
\bibinfo{author}{\bibfnamefont{A.}~\bibnamefont{Weichselbaum}}
  \bibnamefont{and} \bibinfo{author}{\bibfnamefont{J.}~\bibnamefont{{von
  Delft}}} (\bibinfo{year}{2006}), \bibinfo{note}{cond-mat/0607497}.

\bibitem[{\citenamefont{Anders and Schiller}(2005)}]{AS05}
\bibinfo{author}{\bibfnamefont{F.~B.} \bibnamefont{Anders}} \bibnamefont{and}
  \bibinfo{author}{\bibfnamefont{A.}~\bibnamefont{Schiller}},
  \bibinfo{journal}{Phys. Rev. Lett.} \textbf{\bibinfo{volume}{95}},
  \bibinfo{pages}{196801} (\bibinfo{year}{2005}).

\bibitem[{\citenamefont{Bulla}(1999)}]{Bul99}
\bibinfo{author}{\bibfnamefont{R.}~\bibnamefont{Bulla}},
  \bibinfo{journal}{Phys. Rev. Lett.} \textbf{\bibinfo{volume}{83}},
  \bibinfo{pages}{136} (\bibinfo{year}{1999}).

\bibitem[{\citenamefont{Bauer et~al.}(2007)\citenamefont{Bauer, Hewson, and
  Oguri}}]{BHO07}
\bibinfo{author}{\bibfnamefont{J.}~\bibnamefont{Bauer}},
  \bibinfo{author}{\bibfnamefont{A.~C.} \bibnamefont{Hewson}},
  \bibnamefont{and} \bibinfo{author}{\bibfnamefont{A.}~\bibnamefont{Oguri}},
  \bibinfo{journal}{J. Magn. Magn. Mat.} \textbf{\bibinfo{volume}{310}},
  \bibinfo{pages}{1133} (\bibinfo{year}{2007}).

\bibitem[{\citenamefont{Hertz and Edwards}(1972)}]{HE72}
\bibinfo{author}{\bibfnamefont{J.~A.} \bibnamefont{Hertz}} \bibnamefont{and}
  \bibinfo{author}{\bibfnamefont{D.~M.} \bibnamefont{Edwards}},
  \bibinfo{journal}{Phys. Rev. Lett.} \textbf{\bibinfo{volume}{28}},
  \bibinfo{pages}{1334} (\bibinfo{year}{1972}).

\bibitem[{\citenamefont{Spa{\l}ek and Gopalan}(1990)}]{SG90}
\bibinfo{author}{\bibfnamefont{J.}~\bibnamefont{Spa{\l}ek}} \bibnamefont{and}
  \bibinfo{author}{\bibfnamefont{P.}~\bibnamefont{Gopalan}},
  \bibinfo{journal}{Phys. Rev. Lett.} \textbf{\bibinfo{volume}{64}},
  \bibinfo{pages}{2823} (\bibinfo{year}{1990}).

\bibitem[{\citenamefont{Riseborough}(2006)}]{Ris06}
\bibinfo{author}{\bibfnamefont{P.~S.} \bibnamefont{Riseborough}},
  \bibinfo{journal}{Phil. Mag.} \textbf{\bibinfo{volume}{86}},
  \bibinfo{pages}{2581} (\bibinfo{year}{2006}).

\bibitem[{\citenamefont{Zitzler et~al.}(2002)\citenamefont{Zitzler, Pruschke,
  and Bulla}}]{ZPB02}
\bibinfo{author}{\bibfnamefont{R.}~\bibnamefont{Zitzler}},
  \bibinfo{author}{\bibfnamefont{T.}~\bibnamefont{Pruschke}}, \bibnamefont{and}
  \bibinfo{author}{\bibfnamefont{R.}~\bibnamefont{Bulla}},
  \bibinfo{journal}{Eur. Phys. J. B} \textbf{\bibinfo{volume}{27}},
  \bibinfo{pages}{473} (\bibinfo{year}{2002}).

\bibitem[{\citenamefont{Dordevic et~al.}(2006)\citenamefont{Dordevic, Beach,
  Takeda, Wang, Maple, and Basov}}]{DBTWMB06}
\bibinfo{author}{\bibfnamefont{S.~V.} \bibnamefont{Dordevic}},
  \bibinfo{author}{\bibfnamefont{K.~S.~D.} \bibnamefont{Beach}},
  \bibinfo{author}{\bibfnamefont{N.}~\bibnamefont{Takeda}},
  \bibinfo{author}{\bibfnamefont{Y.~J.} \bibnamefont{Wang}},
  \bibinfo{author}{\bibfnamefont{M.~B.} \bibnamefont{Maple}}, \bibnamefont{and}
  \bibinfo{author}{\bibfnamefont{D.~N.} \bibnamefont{Basov}},
  \bibinfo{journal}{Phys. Rev. Lett.} \textbf{\bibinfo{volume}{96}},
  \bibinfo{pages}{017403} (\bibinfo{year}{2006}).

\bibitem[{\citenamefont{Meyer and Nolting}(2001)}]{MN01}
\bibinfo{author}{\bibfnamefont{D.}~\bibnamefont{Meyer}} \bibnamefont{and}
  \bibinfo{author}{\bibfnamefont{W.}~\bibnamefont{Nolting}},
  \bibinfo{journal}{Phys. Rev. B} \textbf{\bibinfo{volume}{64}},
  \bibinfo{pages}{052402} (\bibinfo{year}{2001}).

\bibitem[{\citenamefont{Saso and Itoh}(1996)}]{SI96}
\bibinfo{author}{\bibfnamefont{T.}~\bibnamefont{Saso}} \bibnamefont{and}
  \bibinfo{author}{\bibfnamefont{M.}~\bibnamefont{Itoh}},
  \bibinfo{journal}{Phys. Rev. B} \textbf{\bibinfo{volume}{53}},
  \bibinfo{pages}{6877} (\bibinfo{year}{1996}).

\bibitem[{\citenamefont{Ono}(1998)}]{Ono98}
\bibinfo{author}{\bibfnamefont{Y.}~\bibnamefont{Ono}}, \bibinfo{journal}{J.
  Phys. Soc. Japan} \textbf{\bibinfo{volume}{67}}, \bibinfo{pages}{2197}
  (\bibinfo{year}{1998}).

\bibitem[{\citenamefont{Edwards and Green}(1997)}]{EG97}
\bibinfo{author}{\bibfnamefont{D.}~\bibnamefont{Edwards}} \bibnamefont{and}
  \bibinfo{author}{\bibfnamefont{A.~C.~M.} \bibnamefont{Green}},
  \bibinfo{journal}{Z. Phys. B} \textbf{\bibinfo{volume}{103}},
  \bibinfo{pages}{243} (\bibinfo{year}{1997}).

\bibitem[{\citenamefont{Buu et~al.}(1998)\citenamefont{Buu, Forbes, van
  Steenbergen, Wiegers, Rem\`enyi, Puech, and Wolf}}]{BFSWRPW98}
\bibinfo{author}{\bibfnamefont{O.}~\bibnamefont{Buu}},
  \bibinfo{author}{\bibfnamefont{A.}~\bibnamefont{Forbes}},
  \bibinfo{author}{\bibfnamefont{A.}~\bibnamefont{van Steenbergen}},
  \bibinfo{author}{\bibfnamefont{S.}~\bibnamefont{Wiegers}},
  \bibinfo{author}{\bibfnamefont{G.}~\bibnamefont{Rem\`enyi}},
  \bibinfo{author}{\bibfnamefont{L.}~\bibnamefont{Puech}}, \bibnamefont{and}
  \bibinfo{author}{\bibfnamefont{P.}~\bibnamefont{Wolf}}, \bibinfo{journal}{J.
  Low Temp. Phys.} \textbf{\bibinfo{volume}{110}}, \bibinfo{pages}{311}
  (\bibinfo{year}{1998}).

\end{thebibliography}

\end{document}